\documentclass[aps,pre,twocolumn,superscriptaddress,10pt,longbibliography,floatfix]{revtex4-1}
\usepackage[dvips]{graphicx}
\usepackage{amssymb,amsmath,bbm}
\usepackage{amssymb,amsmath,bbm}
\usepackage[colorlinks]{hyperref}
\usepackage[colorinlistoftodos]{todonotes}

\makeatletter
\def\l@subsubsection#1#2{}
\makeatother

\begin{document}
\title{Importance Sampling of Rare Events in Chaotic Systems}

\author{Jorge C. Leit\~ao}
\affiliation{Max Planck Institute for the Physics of Complex Systems, 01187 Dresden, Germany}
\affiliation{DTU Compute, Technical University of Denmark, Kgs. Lyngby, Denmark}
\author{Jo\~ao M. Viana Parente Lopes}
\affiliation{Department of Physics and Center of Physics, University of Minho, P-4710-057, Braga, Portugal}
\affiliation{Physics Engineering Department, Engineering Faculty of the University of Porto, 4200-465 Porto, Portugal}
\author{Eduardo G. Altmann}
\affiliation{Max Planck Institute for the Physics of Complex Systems, 01187 Dresden, Germany}
\affiliation{School of Mathematics and Statistics, University of Sydney, 2006 NSW, Australia}

\begin{abstract}
Finding and sampling rare trajectories in dynamical systems is a difficult computational task
underlying numerous problems and applications. In this paper we show how to construct Metropolis-Hastings Monte Carlo methods that can efficiently sample rare trajectories in the (extremely rough) phase space of chaotic systems.
As examples of our general framework we compute the distribution of finite-time Lyapunov exponents (in different chaotic maps) and the distribution of escape times (in transient-chaos problems). Our methods sample exponentially rare states in polynomial number of samples (in both low- and high-dimensional systems).
An open-source software that implements our algorithms and reproduces our results can be found in Ref.~\cite{github}.
\end{abstract}
%
\maketitle
\setcounter{tocdepth}{2}
\tableofcontents
\newcommand{\Exp}[1]{\mathbb{E}\left[  #1 \right]}
\newcommand{\Var}[1]{\mathbb{V}\left[  #1 \right]}
\newcommand*\vx{\boldsymbol{x}}
\newcommand*\vh{\boldsymbol{h}}
\newcommand*\vF{\boldsymbol{F}}
\newcommand*\mea{W}  
\newcommand*\p{\pi}
\newcommand*\tstar{t_\star}
\newcommand*\tobs{t_o}
\newcommand*\dos{G}
\newcommand*\uni{U} 
\section{Introduction}
Extreme events play a crucial role in our society.
Landslides, floods, meteorite collisions, solar flares, earthquakes are all events that are rare but often lead to catastrophic consequences to our well being~\cite{Albeverio2006}.
Science often studies extreme events by recreating the process that generates them sufficiently many times.
While in some cases the process can be reproduced experimentally, in many
others (e.g., astronomical and weather events) the best we can do is to simulate physical models.

Physical models often contain non-linearities in the equations of motion that amplify
fluctuations and give origin to extreme events. 
A famous effect of non-linearity is chaos, the extremely sensitive dependence on initial
conditions.
Chaos has a fascinating history dating back to the seminal work of Poincar\'e in 1890~\cite{Holmes1990} and is today a well established field of research with applications in Biology, Geology, Economy, Chemistry, and Physics~\cite{Motter2013,OttBook}.
Chaotic dynamics often hinders our ability to study the evolution of the system analytically, e.g. it forbids describing trajectories in a closed formula.
In this situation, again, often the best we can do is to numerically simulate the model in a computer,
a paradigm popularized by Lorenz since  the 1960s~\cite{Lorenz1963,Motter2013}.  Extreme
events are then studied statistically, over an ensemble of initial conditions. 

In this paper we introduce a framework for performing numerical simulations in
chaotic systems in such a way that rare trajectories are generated more likely than if they would be generated by chance. This is an importance-sampling~\cite{NewmanBarkemaBook,RobertCasellaBook} (or rare-event-simulation~\cite{Bucklew}) strategy that builds on methods that have proven successful in the characterization of rare configurations in various problems. The distinguishing feature of our framework is that it considers general classes of observables~\cite{Lucarini2016} in {\it deterministic} chaotic~\cite{OttBook,LaiTamasBook} systems, being therefore able to find and sample initial conditions leading to extreme events in different problems. 
We apply our framework to two traditional problems in the numerical exploration of chaotic systems:
\begin{itemize}
\item trajectories with high or low finite-time Lyapunov exponents;
\item long-living trajectories in open systems.
\end{itemize}
The results shown in this paper are general, but simulations are done in simple dynamical systems (time-discrete
maps with up to 16 dimensions). Our goal is to illustrate the generic
computational challenges of sampling rare events in chaotic
systems, in line with the tradition that simple systems often possess the basic
mechanisms responsible for extreme events.

The importance of rare trajectories in chaotic systems, including methods designed to find
and sample them, have been studied through different perspectives~\cite{Sweet2001,Bollt2005,LaiTamasBook,Dellago2002,Bolhuis2002,Tailleur2007,Kitajima2011,Laffargue2013a,Patra2015,Wouters2015,Lucarini2016}.
For example, the method \emph{stagger and dagger}, used to find long-living trajectories
of transiently chaotic systems, has been an important tool to characterize chaotic
saddles~\cite{Sweet2001,LaiTamasBook}. \emph{Lyapunov weighted dynamics} is a population Monte Carlo method that was
successfully applied to find trajectories with atypically low or high chaoticity~\cite{Tailleur2007,Philipp2010}. Similar {\it genealogical particle analysis} was applied to compute rare events in simple climate models~\cite{Wouters2015}.
Another powerful and widely used method within trajectory sampling is {\it transition path sampling}~\cite{Dellago2002,Bolhuis2002}, that has been used to sample rare trajectories (e.g. chemical reactions), typically influenced by thermal noise~\cite{Bolhuis2002}.
These are typically trajectories that transit from one stable configuration to another stable configuration~\cite{Bolhuis2002,Grunwald2008}.
These different methods achieve their goal through different, often ingenious, solutions.
Here we aim to construct methods that are applicable to different classes of problems and to quantitatively understand the impact of different parameters and choices on the efficiency of the algorithm. We then show that some of the existing methods can be derived from our construction under suitable approximations.

Our framework relies on Metropolis-Hastings (MH) importance sampling~\cite{NewmanBarkemaBook,RobertCasellaBook}, a well established numerical technique that has been used in statistical physics to study rare events since the 1950s.
MH produces a random walk $\vx \rightarrow \vx'$ in the phase-space of the system
that generates initial conditions leading to extreme events (rare states) more often than they would be found by chance, consequently reducing the computational cost associated with their rareness.
The flexibility of MH is confirmed by its success in numerous fields in Physics, Chemistry, Finance, among many others~\cite{NewmanBarkemaBook,RobertCasellaBook}.
While transition path sampling~\cite{Bolhuis2002}, Lyapunov weighted dynamics~\cite{Tailleur2007}, and genealogical methods~\cite{Wouters2015} already use importance
sampling in deterministic chaotic systems, three fundamental questions remains largely open: 1. can MH be used to systematically sample rare trajectories of (deterministic) chaotic systems? If yes, 2. how and 3. at what (computational) cost?

To answer these questions, we develop a systematic approach to apply MH in 
chaotic systems.
The crucial step is the choice of the proposal distribution of the MH
algorithm, the conditional probability of ``trying'' a state $\vx'$ in the phase-space of
the system, given the current state $\vx$.
The question we have to answer in order for MH to work is: what proposal distribution
guarantees that an observable of the trajectory starting at $\vx'$, e.g. its Lyapunov
exponent, is similar to the same observable of the trajectory starting at $\vx$? 
Our main contribution is a methodology to answer this question for a broad class of
chaotic systems and observables.  More specifically, we show how incorporating properties
of trajectories of chaotic systems in the proposal is a necessary condition to
obtain an efficient MH algorithm. 
This methodology allows to construct efficient MH algorithms to sample rare events in different problems and classes of chaotic systems.
We expect the ideas and formalism presented here to find applications in other problems of chaotic systems and in the study of extreme events more generally.
Therefore, we expect our results to be useful both to those studying extreme events in non-linear systems and to those studying numerical techniques.

The paper is organized as follows: 
  Sec. \ref{sec:chaos} reviews traditional numerical problems in chaotic systems and shows how they can be formulated as a problem of sampling rare trajectories;
Sec. \ref{sec:methods} introduces the MH algorithm as a method to perform importance-sampling simulations in chaotic systems, and shows that naive approaches do not lead to an efficient MH;
  Sec. \ref{sec:framework} shows how to incorporate general features of chaotic systems, such as exponential divergence or self-similarity of some of its properties, in the proposal distribution in order to achieve an efficient MH;
  Sec.~\ref{sec:applications} presents numerical tests of the general framework that
  confirm the applicability of Monte Carlo algorithms to sample rare events in different classes of chaotic systems;
  Sec. \ref{sec:conclusions} summarizes our results and discusses its implications.

\section{Review of  problems}
\label{sec:chaos}\label{sec:stability}

We consider dynamical systems whose states $\vx$ in a phase space $\Omega$, $\vx \in
\Omega \subset \mathbb{R}^D$, evolve in time from an initial condition $\vx=\vx_0$ according to
\begin{equation}
\vx_{t+1} = \vF(\vx_t) = \vF^{t+1}(\vx_0),
\label{eq:evolution}
\end{equation}
where $\vF(\vx) \in \Omega$ and $\vF^t$ is $\vF$ composed $t$ times, $\vF^t(\vx) \equiv \vF(\vF(...\vF(\vx)...))$.
Such discrete-time systems can be obtained from continuous-time systems through a Poincare surface of section, a stroboscopic
projection, or by the time discretization of the numerical integration of differential equations. We are also
interested in the dynamics in the tangent space, which quantifies the divergence between
two initial conditions~\cite{OttBook}.
Specifically, the distance of a state displaced from $\vx$ by $\vh$, $\vx' = \vx + \vh$, to the original state $\vx$ evolves in time according to 
\begin{equation}
\vx'_t-\vx_t = \vF^t(\vx') - \vF^t(\vx).
\end{equation}
Expanding $\vF^t(\vx')$ around $\vF^t(\vx)$ allows $\vx'_t - \vx_t$ to be written as
\begin{equation}
\vx'_t - \vx_t = J_t(\vx)\cdot \vh + \frac{1}{2}\sum_{i=1}^n \sum_{j=1}^n \frac{\partial^2\vF^t(\vx)}{\partial x_i \partial x_j} h_i h_j + O(|\vh|^3)
\label{eq:displacement_all}
\end{equation}
where $J_t(\vx) \equiv {\bf d}\vF^t(\vx)/{\bf d}\vx$ is the Jacobian matrix of $\vF^t$, and $\partial^2\vF^t(\vx)/(\partial x_i \partial x_j)$ is the $(i,j)$ entry of the Hessian matrix of $\vF$.
The first term of Eq.~\ref{eq:displacement_all} can be expanded using the derivative of
the composition and be written as 
\begin{equation}
D(\vx,\vh,t) \equiv
J_t(\vx)\cdot \vh = 
\left(\prod_{i=t-1}^{0}J\left(\vx_{i}\right)\right) \cdot \vh = \vh_t
\label{eq:divergence}
\end{equation}
where $J\equiv J_1$ and $\vh_t$ evolves in the tangent space according to 
\begin{equation}
\vh_0 = \vh \ \ ; \vh_{i+1} = J\left(\vx_{i}\right) \cdot \vh_{i} \ \ .
\label{eq:tangent_evolution}
\end{equation}
For small $|\vh|$, the growth of $\vx'_t - \vx_t$ is characterized by the eigenvalues and eigenvectors of $J^t$~\cite{ChaosBook}.
The largest finite-time Lyapunov exponent (FTLE) of a point $\vx$ can be
defined\footnote{This is sometimes denoted as the stability exponent. See Chapter 4 and 6
  of Ref.~\cite{ChaosBook}.} as
\begin{equation}
\lambda_t(\vx) = \frac{\log(\mu_1)}{t} \ \ ,
\label{eq:finite_time_lambda}
\end{equation}
where $\mu_1$ is the real part of the largest eigenvalue of $J^t$.
Thus, when at least one direction is unstable ($\mu_1>1$), $\vx_t' - \vx_t$ increases exponentially with time, and at most by
\begin{equation}
\vx_t' - \vx_t = \delta_0 e^{\lambda_t(\vx) t} \ \ .
\label{eq:exp_divergence}
\end{equation}
When the system is one dimensional, the ``Jacobian matrix'' is a single number, the product in Eq.~\ref{eq:divergence} is a product of numbers, and the only ``eigenvalue'' is the result of this product. Thus, in this case Eq.~\ref{eq:finite_time_lambda} can be written as 
\begin{equation}
\lambda_t(\vx) =
\frac{1}{t} \sum_{i=0}^{t-1} \log |\frac{dF(\vx_i)}{dx}| \ \ .
\label{eq:finite_time_directional_lambda_sum}
\end{equation}

We now describe two computational problems in chaotic dynamical systems. 
\subsection{Variability of trajectories' chaoticity}
\label{sec:ftle}

For a chaotic system, $\lambda_L \equiv \lambda_{t\rightarrow \infty}>0$ in
Eq.~\ref{eq:finite_time_lambda}. The variation of the (maximum)
{\it finite-time} Lyapunov exponents $\lambda_t$ (FTLE) across different initial conditions
characterizes the variation of chaoticity of the system, yielding a distribution of the
FTLE, $P(\lambda_{\tobs})$, or, equivalently, the distribution of $E \equiv \tobs \lambda_{\tobs}$:
\begin{equation}
P(E) = \int \delta(E - \tobs \lambda_{\tobs}(\vx)) \uni(\vx) {\bf d}\vx \ \ .
\label{eq:lambda_distribution}
\end{equation}
where $\uni(\vx)$ is a chosen probability distribution (e.g. uniform in the phase-space).
The FTLE and its distribution was introduced in the 80s~\cite{Grassberger1985,Grassberger1988,Sepulveda1989} and has been used to study turbulent flows~\cite{Beigie1993}, Hamiltonian dynamics~\cite{Amitrano1992}, chimera states~\cite{Olmi2015}, characterize dynamical trapping~\cite{Grassberger1985,Zaslavsky2002,Szezech2005,Artuso2009,Manchein2012}, among others~\cite{Abarbanel1991,Abarbanel1992}.
The distribution of FTLE is related to the generalized dimensions~\cite{BeckBook} and often follows a large deviation principle, where $\tobs$ is the extensive parameter~\cite{BeckBook,Laffargue2013a}.
In strongly chaotic systems, the distribution of FTLE is Gaussian~\cite{BeckBook}, whereas for intermittent chaos and other weakly chaotic systems the distribution is typically non-Gaussian~\cite{Prasad1999}.

\begin{figure}
\includegraphics[width=\columnwidth]{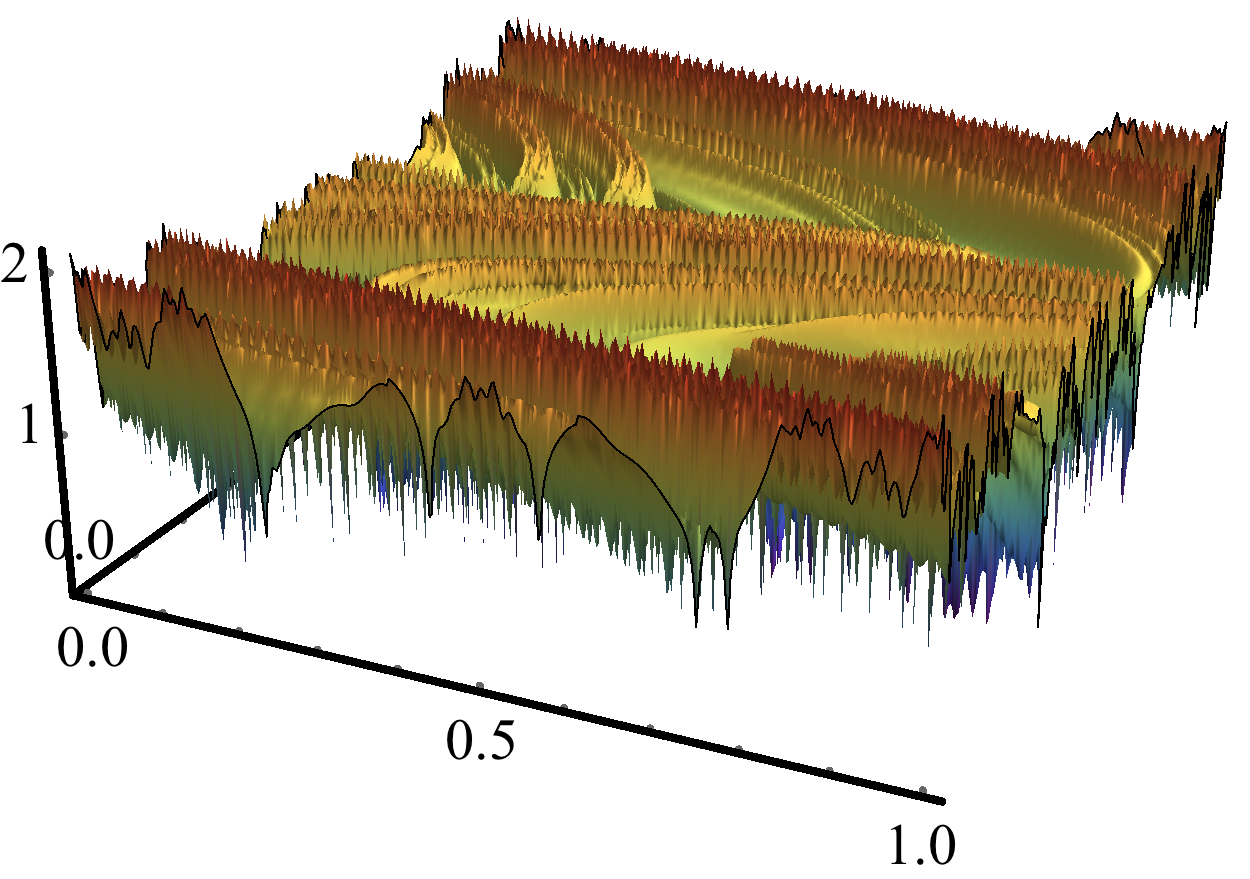}
\includegraphics[width=\columnwidth]{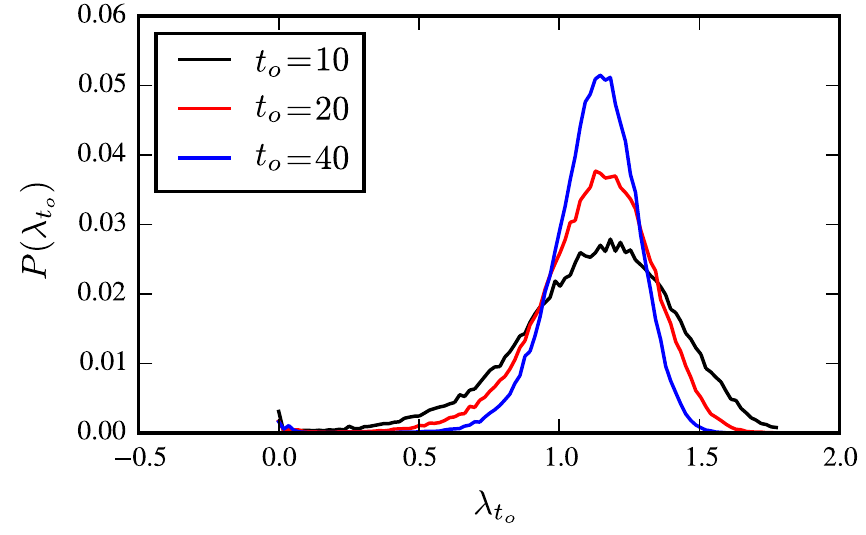}
\caption{
Two main characteristics of the FTLE.
Upper panel: the intricate dependency of $\lambda_{\tobs=4}(\vx)$ (z-axis) with the state $\vx$ (2D, x and y axis).
Lower panel: the distribution of FTLE for different finite times $t$ shows exponential decaying tails with $\lambda_{\tobs}$, and decay with $\tobs$.
The system used was the Standard Map (Eq.~\ref{eq:standard_map}) with $K=6$, over the full phase-space, $\Gamma = \Omega = [0,1]^2$.
$P(\lambda_{\tobs})$ was computed from $10^5$ uniformly distributed initial conditions on
$\Gamma$. 
}
\label{fig:challenges_lyapunov}
\end{figure}

Figure~\ref{fig:challenges_lyapunov} shows the phase-space dependency and
distribution of the FTLE  in one chaotic system. It contains characteristic features
observed in strongly chaotic systems: $\lambda_t$ is a quantity that depends sensitively
on the state $\vx$, and its distribution $P(\lambda_t)$ decays exponentially to zero on
both sides. 
In the limit $\tobs \rightarrow \infty$,  $P(\lambda_{\tobs}) \rightarrow \delta(\lambda_{\tobs} - \lambda_L)$.

The tails of the distribution $P(E)$ play a significant role in the characterization of chaotic systems, as, for example, the higher moments of the distribution are related to higher $q$s in the generalized dimensions $D_q$ of the attractor~\cite{BeckBook}.
Furthermore, the regions of the phase-space with small (large) finite-time Lyapunov exponent are associated with slow (fast) decay of correlations~\cite{Sepulveda1989}, and their characterization has been used to get insight on whether the system is ergodic or not~\cite{Amitrano1992}.
Moreover, trajectories characterized by a low or high finite-time Lyapunov exponent can play a significant role in the dynamics of interfaces in chaotic flows~\cite{Beigie1993} and others~\cite{Tailleur2007}.

A typical analysis of the FTLE is to measure how a quantity, $\mea(\vx)$, depends on the FTLE $\lambda_{\tobs}$~\cite{Sepulveda1989,Amitrano1992,Beigie1993}. This requires estimating an integral of the form 
\begin{equation}
\mea(\lambda_{\tobs}) \equiv \int_\Gamma \mea(\vx) \delta(\lambda_{\tobs} - \lambda_{\tobs}(\vx)) \uni(\vx) {\bf d}\vx
\label{eq:lambda_integral}
\end{equation}
where $\mea(\vx)$ is the pre-selected quantity, $\Gamma$ is a pre-selected sampling region
(often $\Gamma=\Omega$), and $\uni(\vx)$ is the weight attributed to $\vx$ (often uniform, $\uni(\vx) = 1/|\Gamma|$).
Let us look for two examples.
First, consider the problem of estimating the distribution $P(\lambda_{\tobs})$, e.g. Refs.~\cite{Sepulveda1989,Amitrano1992,Prasad1999,Tailleur2007}.
The traditional technique is to sample $M$ states $\vx$ according to $\uni(\vx)=1/|\Gamma|$ and estimate $P(\lambda_{\tobs})$ using the estimator $M_{\lambda}/M$, where $M_{\lambda}$ is the number of samples with $\lambda_{\tobs}(\vx) \in [\lambda_{\tobs}, \lambda_{\tobs} + \Delta \lambda_{\tobs}]$.
Formally, this corresponds to $\mea(\lambda_{\tobs})$ with $\mea(\vx) = 1$. 
The second example is retrieved from Ref.~\cite{Sepulveda1989}.
There, in order to evaluate the contribution of the algebraic region of the phase-space to
the power spectra, the authors decomposed it in two terms corresponding to the power spectra of trajectories with low FTLE and trajectories with high FTLE.
This required estimating the power spectra from a set of trajectories conditioned to an interval of $\lambda$s on the tails of the distribution.
Associating the power spectrum ($S(f)$ in the ref.) with $\mea(\vx) = \mea_f(\vx)$ where $f$ is
the frequency, the power spectra represented in Fig. 4 of Ref.~\cite{Sepulveda1989}
corresponds to integrals (for different frequencies $f$) given by
\begin{equation}
\Exp{\mea_f | \lambda_{\tobs}} = \frac{1}{P(\lambda_{\tobs})}\int_\Gamma \mea_f(\vx) \delta(\lambda_{\tobs} - \lambda_{\tobs}(\vx)) \uni(\vx) {\bf d}\vx \ \ ,
\end{equation}
which contains integrals of the form of Eq.~\ref{eq:lambda_integral}.

Numerically estimating the integral in Eq.~\ref{eq:lambda_integral} is challenging for two reasons: first, $P(\lambda_{\tobs})$ decays with $\tobs$ (lower panel of Fig.~\ref{fig:challenges_lyapunov}):
the distribution of FTLE often follows a large deviation principle,
$P(\lambda_{\tobs}(\vx)) \propto \exp \left(\tobs s(\lambda_{\tobs}) \right)$, where $s$
is intensive in respect to $\tobs$ and is often concave~\cite{BeckBook}.
Consequently, the traditional methodology of sampling states uniformly to find or sample states with increasing
$\tobs$ requires an exponentially high number of initial conditions.
Second, the dependency of $\lambda_{\tobs}(\vx)$  on $\vx$ has multiple local minima and maxima (upper panel of Fig.~\ref{fig:challenges_lyapunov}).
Such rough (fractal~\cite{Richter2016}) landscapes are known to challenge numerical techniques (e.g., simulations get trapped in local minima or maxima)~\cite{Hansmann1993}.

The problem of finding and sampling states with high or low-$\lambda$ has been addressed in the literature with numerical techniques that go beyond traditional uniform sampling~\cite{Tailleur2007,Kitajima2011,Laffargue2013a}.
Such techniques have been successfully applied to find~\cite{Tailleur2007} and sample~\cite{Kitajima2011,Laffargue2013a} states with extremal $\lambda$s in different chaotic systems.
Refs.~\cite{Tailleur2007,Laffargue2013a} use a population Monte Carlo canonical ensemble where $\lambda_t$ plays the role of the energy $E$ to find or sample states with high or low FTLE.
The method computes stochastic trajectories that, from the numerical tests performed, are indistinguishable from (deterministic) trajectories of the system.
Ref.~\cite{Kitajima2011} proposes a flat-histogram simulation to find high or low chaotic states by developing an observable to quantify the chaoticity of the state.

\subsection{Transient chaos}
\label{sec:open_systems}

\newcommand*\exitset{\Lambda}

The best known examples of chaotic systems have a fractal attractor in the phase space~\cite{OttBook}, the Lorenz attractor being the most prominent example~\cite{Lorenz1963}. However, chaotic dynamics can appear also when the fractal invariant set in the phase space is not purely attracting, e.g. it may have stable and unstable directions (a saddle). Trajectories close to this set perform chaotic motion for an arbitrarily long (yet finite) time. This phenomenon appears in a variety of physical systems and is known as transient chaos~\cite{Tel2008,Altmann2013}. 

\begin{figure}[!t]
\centering
\includegraphics[width=\columnwidth]{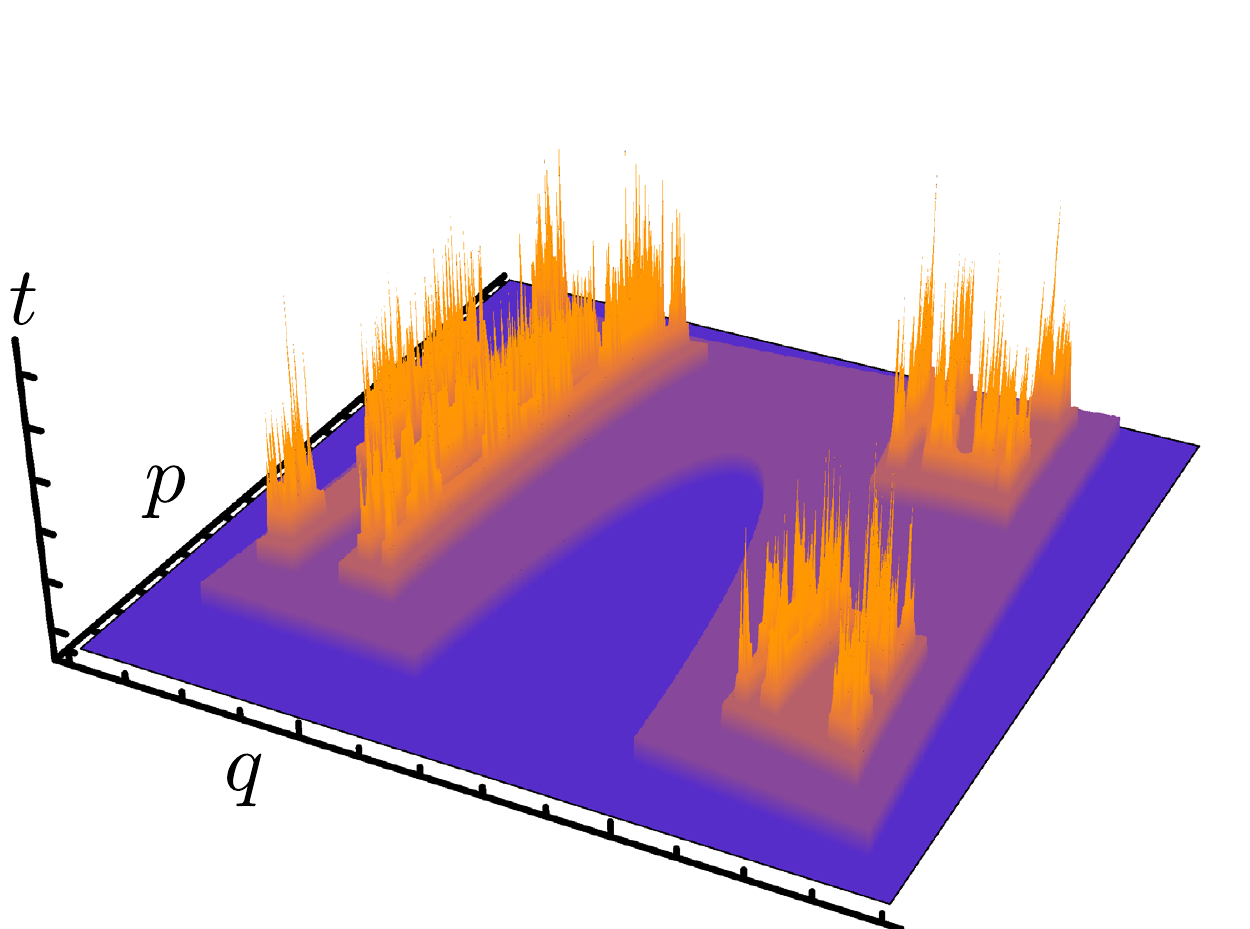}
\includegraphics[width=\columnwidth]{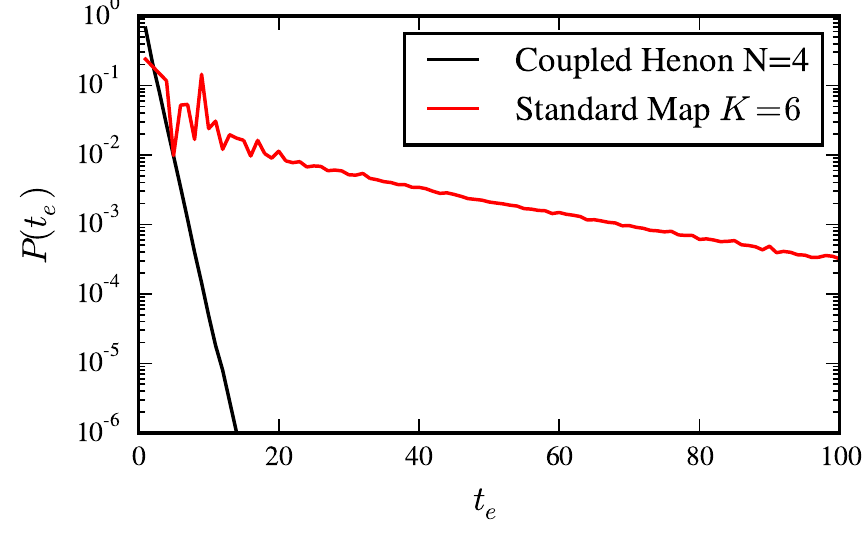}
\caption{
Main characteristics of the escape time of an open chaotic system.
Upper panel: the dependency of $t_e(\vx)$ with the state $\vx$, shows an intricate landscape with multiple local and global maxima. The map is the 4-dimensional coupled H\'enon map (defined in Appendix~\ref{app:maps}), and the two dimensions are a surface of section on the plane $x_2=0, y_2=0$.
Lower panel: the exponential decay of the distribution of escape time of (1) the Coupled H\'enon map with $D=4$ and (2) the Standard Map, both defined in Appendix~\ref{app:maps}. $P(t_e = t)$ was computed by uniformly drawing $10^6$ states $\vx \in \Gamma$, and measure the relative number of times that $t_e(\vx) = t_e$.
}
\label{fig:fractal_landscape}
\end{figure}

Numerical investigations of transiently-chaotic systems are computationally difficult because most trajectories quickly escape the vicinity of the chaotic saddle (on which the chaotic dynamics is properly defined)~\cite{Sweet2001,Bollt2005,Sala2016}. More precisely, the escape time of a state $\vx$ in a pre-selected region $\vx \in \Gamma \subset \Omega$ is defined as the first passage time of its trajectory to a pre-selected exit set $\exitset \subset \Omega$:
\begin{equation}
t_e(\vx) \equiv \min\{ t : \vF^t(\vx) \in \exitset \} \ \ .
\label{eq:te}
\end{equation}
Almost all trajectories start at $\Gamma$ and eventually leave when $\vx_t \in \exitset$, but they do so at different times $t_e$.
Such variability is quantified by the distribution of escape time: the probability that a random initial condition $\vx$ leaves at a given time $t_e$,
\begin{equation}
P(t_e) = \int_\Gamma \delta_{t_e,t_e(\vx)} \uni(\vx){\bf d}\vx \ \ ,
\label{eq:te_distribution}
\end{equation}
where $\delta_{t_e,t_e(\vx)}$ is the Kronecker delta  and $\uni(\vx)$ is the probability density assigned to each state in $\Gamma$, which is often constant, $\uni(\vx) = 1/\int_\Gamma {\bf d}\vx \equiv 1/|\Gamma|$.
Figure~\ref{fig:fractal_landscape} shows the typical features of $t_e(\vx)$: it strongly
depends on the initial state $\vx$ and its distribution $P(t_e)$ decays exponentially to
zero (i.e., it has a constant escape rate $\kappa$).

There are two main numerical techniques to study transiently chaotic systems.
The first is to \emph{find} one long living state $\vx$ with $t_e(\vx) \gg 1$ and to compute an average over states of this trajectory~\cite{LaiTamasBook,Sweet2001}.
For large $t_e(\vx)$, the trajectory between times $[t_e(\vx)/2 - t_s, t_e(\vx)/2 + t_s]$ where $t_s \ll t_e(\vx)/2$ is close to a trajectory on the chaotic saddle.
When the saddle is ergodic, an average over this long trajectory corresponds to an average over the natural invariant density and therefore an average over these states characterizes invariant properties of the system.
The second technique, which we focus in this work, is to compute averages over an ensemble $\uni(\vx)$ of initial conditions $\vx$ that leave the system at time $t_e(\vx) = t_e$~\cite{LaiTamasBook,Altmann2013}.
For small $t_e$ observations depend on the particular initial density $\uni(\vx)$ (a point of interest in itself~\cite{Altmann2013}), while for large $t$ they characterize invariant properties of the system (like in the previous technique, the states $\vF^{t/2}(\vx)$ with $t\rightarrow \infty$ are independent samples of the natural invariant density). 

A typical analysis within sampling transiently chaotic systems is to measure how a quantity, $\mea(\vx)$, changes with increasing escape time $t_e$.
Numerically, this can be written as
\begin{equation}
\mea(t_e) \equiv \int_\Gamma \mea(\vx) \delta_{t_e,t_e(\vx)} \uni(\vx){\bf d}\vx \ \ .
\label{eq:integral_open_systems}
\end{equation}
Let us enumerate 3 examples of computational problems that can be interpreted as numerical estimations of an integral of the form of $\mea(t_e)$ in Eq.~\ref{eq:integral_open_systems}.

\paragraph{Compute the escape time distribution:}
Numerically, $P(t_e)$ is often computed by drawing states from $\uni(\vx)$ (e.g. uniform density), and counting the relative number of states that exited at escape time $t_e$~\cite{LaiTamasBook}.
This corresponds to $\mea(\vx) = 1$ in which case Eq.~\ref{eq:integral_open_systems} reduces to Eq.~\ref{eq:te_distribution}.

\paragraph{Compute generalized dimensions of the chaotic saddle:}
The generalized dimensions are an important property of the chaotic saddle
and its calculation is often performed by box counting~\cite{Dhamala2001,LaiTamasBook,OttBook}.
Essentially, the phase-space is divided in equal, non-overlapping, and space-filling boxes $i=1,...,B(\varepsilon)$ of linear size $\varepsilon$ (intervals in 1 dimension, squares in 2, etc.) and the generalized dimension $D_q$ of exponent $q$ is proportional to $\log \sum_i^B(\varepsilon) \mu_i^q$, where $\mu_i$ is the fraction of points of the saddle that belong to the box $i$~\cite{Dhamala2001}.
Numerically, $\mu_i$ is estimated by first obtaining a set of points $\vx_j$ in the saddle and then counting how many are in the particular box $i$.
Such an estimate can be written as the expectation of an indicator function that tells whether a state in the saddle, $\vF^{t_e/2}(\vx)$ for $t_e(\vx)\gg 0$, is inside the box $i$, $\mea(\vx) = \delta_{\vF^{t_e/2}(\vx) \in i}$.
This expectation can be written as a conditional expectation of $\mea(\vx)$ over states that leave at time $t_e$,
\begin{equation}
\Exp{\mea | t_e} \equiv \frac{1}{P(t_e)}\int_\Gamma \mea(\vx) \delta_{t_e,t_e(\vx)} \uni(\vx){\bf d}\vx \ \ .
\label{eq:expectation_open_systems}
\end{equation}
Computing this essentially requires computing integrals of the form of $\mea(t_e)$ in Eq.~\ref{eq:integral_open_systems}.

\paragraph{Compute the distribution of FTLE on the chaotic saddle:}
The distribution of FTLE $P(\lambda)$ is another important property of the chaotic saddle~\cite{Dhamala2001,LaiTamasBook}.
The FTLE $\lambda=\lambda_t$ for a fixed $t$ of an open system is computed for trajectories on the chaotic saddle. Like in the previous problem, each of these trajectories can be obtained by generating a state $\vx$ according to $\uni(\vx)$ (e.g. uniformly distributed in $\Gamma$) that has a large escape time, $t_e(\vx) \gg 1$, and compute $\lambda_t(\vF^{t_e/2}(\vx))$ using Eq.~\ref{eq:finite_time_lambda} for a fixed $t$.
The distribution of FTLE is then computed from an ensemble of these high-escape-time states by constructing different bins of the histogram $I_\lambda = [\lambda, \lambda + \Delta \lambda]$, and numerically compute the relative number of states in each bin.
Formally, this equates to compute the expected number of states with a given escape time $t_e$ whose FTLE is in a bin, and thus corresponds to
computing a conditional expectation of the form of Eq.~\ref{eq:expectation_open_systems} with $\mea(\vx) = \mea_{\lambda}(\vx) = \delta_{\lambda(\vF^{t_e/2}(\vx)) \in I_{\lambda_t}}$.

\subsection{Summary:  numerical problems in the study of rare events in chaotic systems}
\label{sec:chaos_summary}
\label{SectionRareStates}

To provide an unified treatment of the numerical problems in transient chaos and in computing FTLE of closed systems we use a common notation whenever possible. It indicates also how the methods can be generalized to different problems. Firstly, there is a quantity that we call "observable" and denote by $E(\vx)$:
\begin{itemize}
\item $E(\vx) = \tobs \lambda_{\tobs}(\vx)$ in FTLE of closed systems
\item $E(\vx) = t_e(\vx)$ in strongly chaotic open systems
\end{itemize}
Secondly, we use $\delta(E,E')$ to denote both the Kronecker delta and the Dirac delta ($\delta(E-E')$), for discrete and continuous case respectively. This is needed because, even though in practice we will always consider $E$ to be a discrete function due to binning of the histograms, formally the observable $E$ is either discrete ($t_e$ for maps) or continuous ($\lambda_t$ or $t_e$ for flows).
Thirdly, the observable $E$ is a function of the phase-space of the system but typically depends also on an external quantity $N$ that parameterizes how high $E$ can be:
\begin{itemize}
\item $\tobs$ in FTLE of closed systems
\item (a pre-selected) maximum escape time, $t_{\max}$, in strongly chaotic open systems.
\end{itemize}
This parameter is important to us because, the higher it is, the rarer a state with the maximum or minimum possible $E$ is.
In this notation, the two general problems presented in the previous sections can be summarized as follows:

\begin{itemize}

\item the analysis is conditioned to a projection of $\vx$ into a one-dimensional observable $E(\vx)$.

\item the probability distribution of $E$, $P(E)$, decays exponentially with increasing $E$.

\item the focus of the analysis is on rare states in respect to $P(E)$: states with $E$ in one of the tails of $P(E)$.

\item the rareness increases with $N$.

\end{itemize}
In transient chaos problems we are interested in states with increasing $N=t_{\max}$ that correspond to the tail of $P(E)=P(t_e)$, $P(t_{\max})$. In computations of the FTLE in closed systems we are interested in states with increasing $N=\tobs$ and on the tails of $P(E)$.
These problems share two distinct computational problems: find rare states~\cite{Sweet2001,Bollt2005,LaiTamasBook,Tailleur2007} and sample rare states~\cite{Kitajima2011,Laffargue2013a,Leitao2013,Leitao2014}, which can be formalized as follows:

\begin{itemize}
\item Find rare states: minimize or maximize $E(\vx)$ over a constraining region $\Gamma$, $\vx \in \Gamma \subset \Omega$, for increasing $N$.

\item Sample rare states: compute the integral of a function $\mea(\vx)$ conditioned to a particular value of $E$ and for increasing $N$, over an ensemble of states $\vx$ distributed according to $\uni(\vx)$ in a constraining region of the phase-space $\vx \in \Gamma \subset \Omega$:
\begin{equation}
\mea(E) = \int_\Gamma \mea(\vx) \delta(E,E(\vx)) \uni(\vx) {\bf d}\vx \ \ .
\label{eq:integral}
\end{equation}
\end{itemize}

The numerical challenges of these two numerical problems are common: the states $\vx$ are exponentially difficult to find with increasing $N$ (or $E$), the function $E(\vx)$ contains multiple local and global minima embedded in fractal-like structures, and a potential high dimensionality $D$ of the phase-space (see Figs.~\ref{fig:challenges_lyapunov} and~\ref{fig:fractal_landscape}).
The goal of this paper is to develop a systematic approach to tackle the two numerical problems and these three challenges.


\section{Review of methods}
\label{sec:methods}

\subsection{Importance Sampling}
\label{sec:importance_sampling}
As described in the previous section, the traditional methodology to compute an integral of the form of Eq.~\ref{eq:integral} is to draw $m$ samples $\{ \vx_i \}$ distributed according to $\uni(\vx)$ from the relevant region $\Gamma$ and approximate the integral $\mea(E)$ by the estimator
\begin{equation}
\overline{\mea(E)} \equiv \frac{1}{m}\sum_{i=1}^{m} \delta(E, E(\vx)) \mea(\vx_i)
\label{eq:simple_estimator}
\end{equation}
where $\delta(E, E(\vx)) = 1$ when $E(\vx_i)\in[E, E+\Delta E[$ and zero otherwise (when E is discrete, $\Delta E = 1$).
The relative distance of the estimator $\overline{\mea(E)}$ to $\Exp{\mea(E)}$ is quantified by the ratio $\epsilon(E) \equiv \sigma\left[ \overline{\mea(E)} \right]/\Exp{\mea(E)}$ where $\sigma \left[ \overline{\mea(E)} \right]^2 \equiv \Exp{\overline{\mea(E)}^2} - \Exp{\overline{\mea(E)}}^2$.
For the estimator in Eq.~\ref{eq:simple_estimator}, this is given by
\begin{equation}
\epsilon(E) \propto \frac{1}{\sqrt{m \dos(E)}} \ \ .
\label{eq:variance}
\end{equation}
where $\dos(E)$ is the density of states: the number of states per bin with an observable $E$.\footnote{$G(E) = P(E)$ when each state is equally weighted, $\uni(\vx) = 1/|\Gamma|$.}
Therefore, the number of samples $m_*$ required to achieve a given precision $\epsilon_*$ for a given $E_*$ is $m_*(E_*) \propto 1/G(E_*)$.
The critical problem in sampling rare states is that, because $G(E)$ decays exponentially with $E$, $m_*(E)$ increases exponentially with $E$.

Importance sampling techniques aim to improve this scaling by drawing samples from a distribution $\p(\vx) \neq \uni(\vx)$ on the phase-space (e.g. non-uniformly in the phase-space)~\cite{RobertCasellaBook}.
Specifically, consider $m$ independent samples $\{ \vx_i \}$ drawn from $\p(\vx)$, and the function $\p(\vx)$ to depend only on $E$, $\p(\vx) = \p(E(\vx)) = \p(E)$.
Because the samples are not drawn from $\uni(\vx)$, the estimator in Eq.~\ref{eq:simple_estimator} would be biased.
Importance sampling uses an unbiased estimator for $\mea(E)$ given by~\cite{RobertCasellaBook}
\begin{equation}
\overline{\mea(E)} \equiv \frac{1}{m}\sum_{i=1}^{m} \delta(E,E(\vx_i))\mea(\vx_i)\frac{\uni(\vx_i)}{\p(\vx_i)} \ \ ,
\label{eq:nonuniform_estimator}
\end{equation}
which reduces to Eq.~\ref{eq:simple_estimator} when $\p(\vx)=\uni(\vx)$.
The advantage of importance sampling is that the relative error of the estimator in Eq.~\ref{eq:nonuniform_estimator} is given by
\begin{equation}
\epsilon(E) \propto \frac{1}{\sqrt{m \dos(E) \p(E)}} \ \ .
\label{eq:variance2}
\end{equation}
This is because, when sampling from $\p(\vx) = \p(E(\vx))$, the expected number of samples with a given $E$, $m(E)$, is equal to
\begin{equation}
m(E) = m \dos(E)\p(E) \ \ .
\label{eq:m_samples}
\end{equation}
Equation~\ref{eq:variance2} implies that the function $\p(E)$ can be chosen to favour states $\vx$ with observable $E$ on the tails of $G(E)$ and therefore improve the precision of the estimator on these tails.

The standard deviations in Eqs.~\ref{eq:variance},\ref{eq:variance2} were obtained assuming that the $m$ samples were independent.
In traditional methodologies such as uniform sampling, this is the case.
However, in the algorithms discussed below, it is not.
Therefore, it is necessary to modify Eq.~\ref{eq:variance2} for the case where the samples $\{\vx_i\}$ are drawn from $\p(\vx)$ and are also correlated.
This modification is given by
\begin{equation}
\epsilon(E) \propto \sqrt{\frac{1 + 2T(E)}{m \dos(E) \p(E)}} \ \ ,
\label{eq:variance3}
\end{equation}
where $T(E)$ is the autocorrelation time~\cite{RobertCasellaBook}, which increases with the correlation of the samples.

Efficient importance-sampling~\cite{NewmanBarkemaBook,RobertCasellaBook,Bucklew} techniques have to address the following three steps: 
\begin{enumerate}
\item choose a suitable $\p(\vx)$
\item have a method to generate samples from $\p(\vx)$
\item minimize the autocorrelation time $T$
\end{enumerate}
A defining point in our method is our choice for the Metropolis-Hasting algorithm to address point 2, differently from Refs.~\cite{Tailleur2007,Wouters2015} which address similar problems through a different choice for point 2 (cloning trajectories). Below we first discuss point 2, then point 1, and finally point 3. 

\subsection{Metropolis-Hastings algorithm}
\label{alg:metropolis}

The Metropolis-Hastings (MH) algorithm asymptotically generates states $\vx$ according to $\p(\vx)$ using a
a Markovian, ergodic and detailed balance random walk in the sampling region $\Gamma$~\cite{RobertCasellaBook}.
This random walk is initialized from a random state $\vx \in\Gamma$ and evolves to a new state $\vx' \in \Gamma$ with a transition probability $P(\vx'|\vx)$ chosen such that asymptotically the states $\vx$ are drawn according to $\p(\vx)$.
In Metropolis-Hastings, $P(\vx'|\vx)$ is written as
$$P(\vx'|\vx)=g(\vx'|\vx)a(\vx'|\vx),$$
where  $g(\vx'|\vx)$ is the (proposal) distribution used to generate new states and
$a(\vx'|\vx)$ is the (acceptance) distribution used to select them.
The random walk fulfills detailed balance because the acceptance probability is chosen as~\cite{RobertCasellaBook}
\begin{equation}
a(\vx'|\vx) = \min \left( 1, \frac{g(\vx|\vx')}{g(\vx'|\vx)}   \frac{\p(\vx')}{\p(\vx)}    \right) \ \ .
\label{eq:acceptance}
\end{equation}
Metropolis-Hastings algorithm is not the only way to achieve this.
Another popular and alternative method to sample from $\p(\vx)$ is population Monte Carlo, which instead of a random walk, uses multiple stochastic trajectories that are cloned and destroyed. See e.g. ref.~\cite{Laffargue2013a} for an application to the problem of the FTLE described above.

Algorithmically, the MH algorithm is implemented as follows: choose a region $\Gamma$ and a random initial condition $\vx \in \Gamma$.
Evolve the random walk in time according to:
\begin{enumerate}
\item Propose a state $\vx'$ drawn from $g(\vx'|\vx)$;
\item Compute $a(\vx'|\vx)$ replacing $\vx$ and $\vx'$ in Eq.~\ref{eq:acceptance};
\item Generate a random number $r$ in $[0,1]$. If $r<a(\vx'|\vx)$, make $\vx'$ to be the new $\vx$;
\item Store $\vx$ and go to 1.
\end{enumerate}
The set of sub-steps 1-4 brings the random walk from its current state $\vx$ to the next state and it is called a Markov step.
After a transient number of steps where the algorithm converges to the asymptotic distribution, the stored states $\vx$ are (correlated) samples drawn from $\p(\vx)$, and can be directly used in Eq.~\ref{eq:nonuniform_estimator}.


\subsection{Sampling distribution}

\paragraph{Canonical ensemble}
\label{sec:canonical_ensemble}

A sampling distribution $\p(\vx)$ often used is the canonical distribution~\cite{NewmanBarkemaBook}
\begin{equation}
\p(\vx) = \p(E(\vx)) \propto e^{-\beta E(\vx)} \ \ .
\label{eq:canonical_ensemble}
\end{equation}
In the context of rare states, the canonical ensemble is useful because the number of sampled states $m(E)$ in Eq.~\ref{eq:m_samples} becomes
\begin{equation}
m(E) \propto mG(E) e^{-\beta E} \propto m e^{-\beta E + S(E)} \ \ .
\label{eq:canonic_histogram}
\end{equation}
In particular, the maximum of $m$ is at $E^*$ solution of $\beta = dS/dE(E^*)$.
Therefore, $\beta$ tunes which value of the observable $E$ is sampled the most.
For example, the Lyapunov weighted dynamics in Ref.~\cite{Laffargue2013a} uses this distribution (Eq. 15 of the ref. with $\alpha$ replaced by $\beta$).

\paragraph{Flat-histogram ensemble}
\label{sec:flat-histogram}
\label{sec:wang-landau}

Another distribution often used in the literature of Metropolis-Hastings~\cite{Torrie1977,Swendsen1986,Marinari1992,Sugita1999,Lee1993,Berg1991,Wang2001,Lopes2006} is the flat-histogram (or multicanonical), given by~\cite{Berg1991,Wang2001}
\begin{equation}
\p(\vx) = \p(E(\vx)) \propto \frac{1}{G(E(\vx))}, \ \ E \in [E_{\min}, E_{\max}],
\label{eq:flathistogram_ensemble}
\end{equation}
for a given choice of $E_{\min}, E_{\max}$ that defines the region of interest on the observable $E$.
This is known as flat-histogram because, replacing Eq.~\ref{eq:flathistogram_ensemble} in Eq.~\ref{eq:m_samples} leads to a constant average number of samples on each $E$,
\begin{equation}
m(E) = \text{const.}
\label{eq:constant_histogram}
\end{equation}
Consequently, the dependence of the variance in Eq.~\ref{eq:variance3} is only due to the autocorrelation $T(E)$, which implies that the computational cost to draw a state on the tail of $G(E)$ is no longer driven by the exponential decrease of $G(E)$, but by the computational cost to draw uncorrelated samples from $\p(\vx)$.


The main limitation of the flat-histogram is that it requires knowing $G(E)$ in advance, which is very often unknown.
The most well known, that we use here, is the Wang-Landau algorithm~\cite{Wang2001}, that modifies the Metropolis-Hastings algorithm to a non-markovian chain that asymptotically converges to a flat-histogram Metropolis-Hastings.
The Wang-Landau algorithm starts with an approximation of $G(E)$, $G_{WL}(E) = 1/ |E_{\max}-E_{\min}|$, and, on
step 4 of the MH (see algorithm in Sec.~\ref{alg:metropolis}), it multiplies $G_{WL}(E(\vx))$ ($\vx$ is the current state of the random walk) by a constant $f>1$.
After a given number of steps, $f$ is reduced by a factor 2 and this procedure is repeated until a final $f_{\min} \simeq 1$ is reached~\cite{Wang2001}.
The value $f_{\min}$ and how $f$ is reduced dictates how close $G_{WL}(E)$ will be from $G(E)$~\cite{Wang2001,Zhou2005,Belardinelli2008a}.

\subsection{Characterisation of the efficiency}
\label{sec:efficiency}

The relative error $\epsilon(E)$ depends on the value of $N$ and on the particular value of $E/N$.
To avoid discussing the dependency on $E/N$, the efficiency of the flat-histogram is often quantified in terms of the average round-trip~\cite{Trebst2004,Dayal2004,Lopes.phd2006}, which is an upper bound for the number of Markov steps $m$ (samples) required to obtain an uncorrelated sample from $\p(\vx)$~\cite{Costa2005,Lopes.phd2006}.
The round-trip, $\tau$, is the average number of steps required for the algorithm to go from a state $\vx$ with $E(\vx)=E_{\min}$ to a state $\vx$ with $E(\vx)=E_{\max}$ and return back.~\footnote{Numerically, the round-trip time is computed by having a boolean value tracking whether the random walk is moving in the direction $E_{\min} \rightarrow E_{\max}$ or $E_{\max} \rightarrow E_{\min}$, and use it to measure the total number of steps required to make the full round-trip.}
Ref.~\cite{Leitao2013} shows how the autocorrelation time of a canonical ensemble is related to the autocorrelation time of a flat-histogram, and therefore we will use here the round-trip time of a flat-histogram simulation to quantify the efficiency for both distributions.
The uniform sampling has a round-trip that increases exponentially with $N$: on average it takes $1/G(E_{\max})$ samples to get one sample with $E_{\max}$, and $1/G(E_{\max})$ increases exponentially with $N$.

Importance sampling Monte Carlo is widely used in statistical physics because the computational cost often scales polynomially with $N$~\cite{Lopes.phd2006,RobertCasellaBook}, which is a dramatic improvement over uniform sampling.
Under the hypothesis that 
(a) $\Delta E \equiv E(\vx') - E(\vx) \approx 1 \ll N$ and
(b) the correlation between the different $E(\vx)$ of the random walk decay fast enough, it can be shown that the roundtrip $\tau$ scales as~\cite{Lopes.phd2006,Leitao2014} 
\begin{equation}
\tau(N) \sim N^2 \ \ .
\label{eq:round_trip_scaling}
\end{equation}
For example, consider the problem of sampling states of an open chaotic system with different escape times $t_e$ from $t_e = 1$ up to $t_e = t_e^{\max}$.
In this case, $E_{\min} = 1$ and $E_{\max} = N = t_e^{\max}$. Thus, under the hypothesis (a) and (b) above, the round-trip is expected to scale as 
\begin{equation}
\tau(t_e^{\max}) \sim (t_e^{\max})^2 \ \ .
\end{equation}
There are known deviations of the scaling in Eq.~\ref{eq:round_trip_scaling} leading to a higher exponent, $N^{2+z}$ with $z>0$~\cite{Trebst2004,Dayal2004,Lopes.phd2006,Leitao2014} and there are two common situations where this happens: (1) the acceptance rate decreases drastically with $N$ (hypothesis (a) above is violated); (2) autocorrelation drastically increases with $N$ (hypothesis (b) above is violated)~\cite{Lopes.phd2006,Leitao2014}.
Nevertheless, these do not qualitatively change the argument: Monte Carlo with flat-histogram has the potential of generating rare states polynomially with increasing $N$, while uniform sampling generates rare states exponentially with increasing $N$.


\subsection{Summary: challenges for the application of Monte Carlo methods to chaotic systems}
\label{sec:mc_challenge}

To achieve a MH algorithm that scales polynomially with N, the autocorrelation time needs to be low.
A key ingredient for an efficient algorithm is a good proposal distribution $g(\vx'|\vx)$~\cite{Roberts1997}.
The ideal proposal distribution of a Metropolis-Hastings algorithm draws $\vx'$ independently of $\vx$ according to $\p(\vx')$, $g(\vx'|\vx) = \p(\vx')$.
This is because i) the acceptance in Eq.~\ref{eq:acceptance} is always one and ii) each step of the random walk generates an independent sample $\vx$, which implies that the error in Eq.~\ref{eq:variance3} is minimal. 
The difficulty of sampling rare events in chaotic systems is that a useful $\p(\vx)$ to sample them is a difficult function to sample from.
For concreteness, consider the problem of sampling high escape times in the open tent map defined in Appendix~\ref{sssec.tent} -- Eq.~\ref{eq:open_tent}, whose $t_e(\vx)$ is represented in Fig.~\ref{fig:open_tent} -- and consider the canonical sampling distribution, $\p(\vx) \propto \exp(-\beta t_e(\vx))$.
In this example, $\p(\vx) \propto \exp(-\beta 1)$ between $[1/a, 1-1/a]$ and so forth ($\exp(-\beta t_e)$) in subsequent intervals.
The number of intervals increases as $2^{t_e}$.
Therefore, sampling from $\p(\vx)$ would require enumerating every interval, sample one at random according to a correct distribution that depends on $\beta$, and then sample a uniform point within that interval.
While this can be done in simple maps such as the open tent map, this is unfeasible in a general system where the $t_e(\vx)$ dependency is unknown.

One way to approach the problem could be to consider $g(\vx'|\vx)$ to be the uniform distribution over $\Gamma$.
One could imagine that changing the sampling distribution $\p(\vx)$ could decrease the variance of the estimator, since this gives more preference to rarer states.
However, this is not the case: changing the sampling distribution alone does not decrease the scaling of the variance of the estimator. 
This is because changing the sampling distribution (e.g. using a canonical ensemble) leads to an exponential increase of the autocorrelation time $T(E)$ with $E$, see Appendix \ref{app:acceptance_derivation}, making it as efficient as traditional uniform sampling.

In summary, Metropolis-Hastings is an excellent candidate to approach the numerical challenges found in the study of rare events in chaotic systems.
Firstly, because it is grounded in strong mathematical results such as importance sampling theorem and asymptotic convergence of Markov processes.
Secondly, because it is formulated with very little assumptions about the system, the observable of interest or the dynamics of the system, which gives enough freedom to adapt it to the specific aim (sampling or finding), observable, and system.
Thirdly, because there seems to be no theoretical reason for the sampling to be exponential; Metropolis-Hastings is used to sample rare states in polynomial time in other problems of statistical physics.
Finally, because the numerical problems found in chaotic systems can be re-written as problems where Metropolis-Hastings is suitable for.
On the other hand, the optimal proposal of Metropolis-Hastings is unfeasible in chaotic systems, and without any extra information about the system, Metropolis-Hastings is as efficient as the traditional uniform sampling.


\section{Proposal distribution}
\label{sec:framework}
The problem we address in this section is: how to incorporate general properties of chaotic systems into the proposal distribution in such a way that the Metropolis-Hastings algorithm becomes efficient? We first set an aim for the proposal distribution and we then show how this aim can be achieved in the different problems involving deterministic chaotic system.

\subsection{Aim of the proposal distribution}
\label{sec:derivation_ratio}

\newcommand*\astar{a_\star}

The goal of the proposal distribution $g(\vx'|\vx)$ we construct here will be to bound the acceptance rate in Eq.~\ref{eq:acceptance} away from 0 and 1.
Since $a(\vx'|\vx)$ in Eq.~\ref{eq:acceptance} depends on $\vx'$, it is essential to look at its expectation over the proposal,
\begin{equation}
\Exp{a(\vx'|\vx)|\vx} \equiv \int_\Gamma a(\vx'|\vx) g(\vx'|\vx) {\bf d}\vx' \ \ ,
\label{eq:expectation_x_prime}
\end{equation}
Our goal is to construct a proposal distribution such that\footnote{E.g. Ref.~\cite{Roberts1997} computed $\astar = 0.234$.}
\begin{equation}
\Exp{a(\vx'|\vx)|\vx} = \astar \ \ .
\label{eq:bounded_acceptance}
\end{equation}
The motivation to set this as the starting point (and cornerstone) of our method is that it avoids the two typical origins of high correlation times $T$.
When the acceptance is low, $T$ increases because the method remains stuck in the same state $\vx$ for long times.
High acceptance typically indicates that $\vx'$ is too close to $\vx$ (often $E(\vx') = E(\vx)$), which implies that the simulation moves
too slowly (in $E$ and in $\Omega)$.
The goal here is not that the acceptance is exactly $\astar$, but rather that it remains bounded from the extremes and that it does not strongly depend on $N$.

In general it is non-trivial to construct a proposal distribution that guarantees a constant acceptance.
Thus, the next step is to approximate Eq.~\ref{eq:bounded_acceptance} by a simpler condition.
Let us first notice that $\p(\vx)$ only depends on $\vx$ through $E_{\vx} \equiv E(\vx)$, $\p(\vx) = \p(E_{\vx})$.
Therefore, at the very least, the proposal $g(\vx'|\vx)$ should guarantee that $\vx'$ is generated from $\vx$ in such a way that $\p(E_{\vx'})$ is neither too close (high acceptance) nor too far (low acceptance) from $\p(E_{\vx})$.
Quantitatively, this can be written as
\begin{equation}
\Exp{\frac{\p(E_{\vx'})}{\p(E_{\vx})}|\vx} = a
\label{eq:base_energy_condition}
\end{equation}
where $0<a<1$ is a constant.
When the proposal is able to achieve a small variation of $E$, $\p(E_{\vx'})$ can be expanded in Taylor series around $E_{\vx'} = E_{\vx}$, which allows to write 
\begin{equation}
\frac{\p(E_{\vx'})}{\p(E_{\vx})} = 1 + \frac{d\log \p(E_{\vx})}{dE} (E_{\vx'} - E_{\vx}) \ \ .
\label{eq:taylor_expansion}
\end{equation}
\footnote{In statistical physics an heuristics often used in Metropolis-Hastings is $\Delta E \equiv E_{\vx'} - E_{\vx} \approx 1$. E.g. it is used in the derivation of Eq.~\ref{eq:round_trip_scaling}, and has been used for example in spin systems (single spin flip)~\cite{Lopes.phd2006}, ensemble of complex networks (single link exchange)~\cite{Fischer2015}, and proteins~\cite{Grassberger1997}.}
Inserting Eq.~\ref{eq:taylor_expansion} in Eq.~\ref{eq:base_energy_condition}, an average constant acceptance is thus achieved when
\begin{equation}
\Exp{ E_{\vx'} - E_{\vx} | \vx } = \frac{a - 1}{d\log \p(E_{\vx})/dE} \ \ .
\label{eq:energy_condition}
\end{equation}
This equation, the main result of this section, is a condition that an efficient Metropolis-Hastings imposes to the proposal distribution in terms of the average difference in the observable $E$.
This condition is non-trivial because it depends on the particular $\pi$, $E$, $\vF$, and $\vx$.
For the sampling distributions discussed in the previous section, we obtain:

\paragraph{Canonical ensemble:}
when $\p(\vx) = e^{-\beta E_{\vx}}$, the condition in Eq.~\ref{eq:energy_condition} is given by 
\begin{equation}
\Exp{ E_{\vx'} - E_{\vx} | \vx } = \frac{1 - a}{\beta} \ \ .
\label{eq:canonical_energy_condition}
\end{equation}
That is, the higher the $\beta$, the closer the proposed $E_{\vx'}$ has to be from $E_{\vx}$.

\paragraph{Flat-histogram}
When $\p(\vx) \propto 1/G(E_{\vx})$, the condition in Eq.~\ref{eq:energy_condition} is given by 
\begin{equation}
\Exp{ E_{\vx'} - E_{\vx} | \vx } = \frac{1 - a}{\frac{d\log G}{dE}(E_{\vx})} \ \ .
\label{eq:flat_energy_condition}
\end{equation}
When $E_{\vx}$ is close to the maximum of $G(E)$, the derivative of $\log G$ approaches 0 and $E_{\vx'}$ can be arbitrary distant from $E_{\vx}$.
As $E_{\vx}$ deviates from the maximum of $G(E)$, smaller changes are necessary to achieve
a constant acceptance.
Note that equation~\ref{eq:energy_condition} is valid for the Metropolis-Hastings algorithm in general and should be of interest also in other contexts
(e.g., arbitrarily large number of spins can be flipped close to the maximum of the density of states).

\subsection{Propose correlated trajectories}
\label{sec:correlation}
\label{sec:tstar}

The proposal distribution requires correlating the trajectory starting at $\vx'$ with a trajectory starting at $\vx$ such that Eq.~\ref{eq:energy_condition} holds.
Fulfilling this requirement requires the ability to control $\Exp{ E_{\vx'} - E_{\vx} | \vx }$, which requires a procedure to correlate the state $\vx'$ with $\vx$.
The aim of this section is to introduce a quantification of the correlation of two trajectories of finite-time $\tobs$ that can be related with $\Exp{ E_{\vx'} - E_{\vx} | \vx }$, and present two different proposal distributions that propose a state $\vx'$ on which this correlation is controlled.

\newcommand*\tshift{{t_{\text{shift}}}}

\begin{figure}[!t]
\centering
\includegraphics[width=\columnwidth]{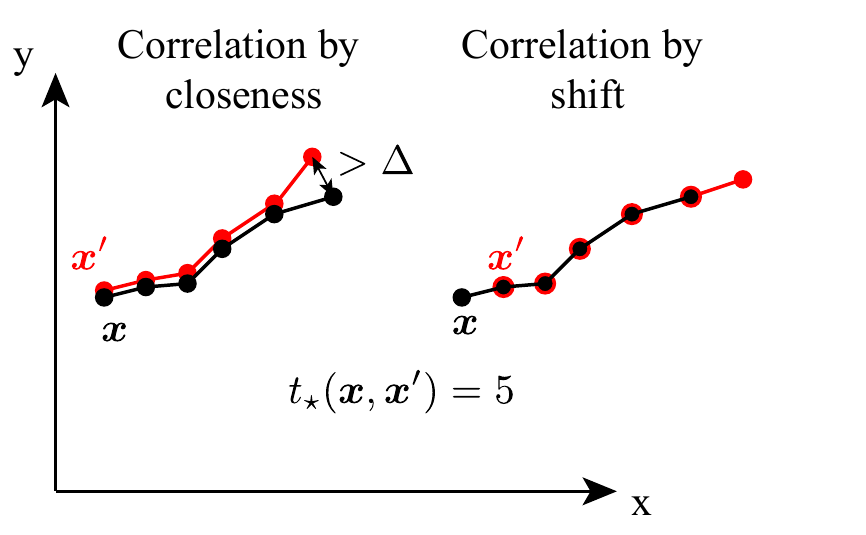}
\caption{
Two trajectories of length $\tobs=6$ starting at $\vx$ and $\vx'$ can be correlated for five steps ($\tstar(\vx,\vx')=5$) by two different mechanisms:
(left panel) they start close from each other and are indistinguishable within $\Delta$ up to time $\tstar$;
(right panel) they start shifted (by $\tshift=1$ here) from each other and are thus indistinguishable for $\tobs - \tshift = \tstar = 5$ steps.
}
\label{fig:tstar}
\end{figure}

The observables $E$ introduced in section~\ref{sec:chaos_summary} are all dependent not only of
$\vx$, but also of the full trajectory of length $\tobs$ starting $\vx$.\footnote{The
  escape time $t_e$ requires that $\vF^t(\vx) \notin \Lambda \ \ , \forall t$; the FTLE
  $\lambda_{\tobs}$ is a sum of terms computed over the trajectory.} 
One natural way to quantify the similarity of two trajectories is the length the two
trajectories remain within a distance $\Delta$ much smaller than a characteristic length of $\Gamma$.
Formally, this can be quantified by $\tstar(\vx,\vx') = \max \{ t \leq \tobs : |\vF^{t}(\vx) - \vF^{t}(\vx')| \leq \Delta \}$.
Under this definition, $0 \leq \tstar(\vx,\vx') \leq \tobs$. When $\vx' = \vx$, $\tstar(\vx,\vx') = \tobs$ because the trajectories are the same; when $\vx'$ is far from $\vx$, $\tstar(\vx,\vx') = 0$.
However, there is another possibility for two trajectories starting at $\vx'$ and $\vx$ to be similar: when the two trajectories are similar apart from a shift in time, see right panel of figure~\ref{fig:tstar}.
One way to include both cases in our measure of similarity is to define $\tstar(\vx,\vx')$ as
%
\begin{equation}
\tstar(\vx,\vx') \equiv \max \{ t \leq \tobs - \tshift : |\vF^{t}(\vx') - \vF^{\tshift + t}(\vx)| \leq \Delta \}.
\label{eq:tstar}
\end{equation}
For $\tshift=0$, this recovers the case of two trajectories starting close to each other; $\tshift \neq 0$, it includes situations where a trajectory starts at $\vx'$ close to $\vF^{\tshift}(\vx)$.
This definition is motivated by the concept of symbolic sequences.~\cite{OttBook}
The similarity of the trajectory starting from $\vx'$ with the one starting from $\vx$ can be quantified by the number  of symbols that both trajectories share, which corresponds to the $\tstar(\vx,\vx')$ in Eq.~\ref{eq:tstar}.
The definition in Eq.~\ref{eq:tstar} avoids the necessity of the existence of a phase-space partition, but, for the purposes of the argument below, the two trajectories share a sequence of $\tstar$ states that are close within $\Delta$.

The average correlation between two states whose one is drawn according to a proposal distribution is here defined by
\begin{equation}
\tstar(\vx) \equiv \Exp{\tstar(\vx,\vx')|\vx} = \int_\Gamma g(\vx'|\vx) \tstar(\vx,\vx') {\bf d}\vx' \ \ .
\end{equation}
Notice that this quantity does not depend on the particular sampling distribution or algorithm; it is a function of the proposal distribution and the state $\vx$.
The goal of the next sub-sections is to construct proposal distributions that guarantee a given average correlation $\tstar(\vx)$.
We will show, for example, that we can enforce an average correlation $\tstar(\vx)$ if we use a normal distribution centered around $\vx$ with a specific standard deviation as our proposal distribution.

\subsubsection{Shift proposal}
\label{sec:shift_proposal}

One proposal that guarantees that trajectories are correlated by $\tstar$ is the shift proposal, originally introduced in Ref.~\cite{Dellago2002} in the context of sampling paths of chemical reactions, and has also been used in Ref~\cite{Grunwald2008}. It consists in proposing a state $\vx'$ that is a forward or backward iteration of $\vx$, $\vx' = \vF^\tshift(\vx)$, where $\tshift$ is a free parameter.
The relation between $\tshift=\tshift(\vx)$ and $\tstar=\tstar(\vx)$ is that a shift of $\pm \tshift$ guarantees that $\tobs - \tshift$ elements of the original trajectory are preserved.
Therefore, this proposal guarantees that $\tstar$ elements are preserved when $|\tshift| = \tobs - \tstar$, see right panel of Fig.~\ref{fig:tstar}.
Because detailed balance has to be enforceable, the proposal must contain backward and forward shifts.
A proposal that automatically fulfils detailed balance is one on which the backward and forward shifts are equally likely:
\begin{equation}
g(\vx'|\vx) = \frac{1}{2}\delta \left( \vx' - \vF^\tshift(\vx) \right) + \frac{1}{2}\delta \left( \vx' - \vF^{-\tshift}(\vx) \right),
\label{eq:shift_proposal}
\end{equation}
with $\tshift = \tshift(\vx) = \tobs - \tstar(\vx)$.
Given the target average correlation $\tstar(\vx)$, this proposal can be implemented as follows: generate a random number $r\in[0,1]$; if $r<0.5$, make $\vx' = F^{\tobs - \tstar(\vx)}(\vx)$, else, make $\vx' = F^{\tstar(\vx) - \tobs}(\vx)$.

This proposal unfortunately has some disadvantages:
i) a priori there is no guarantee that $\vF^\tshift(\vx) \in \Gamma$. It is applicable when $\Gamma = \Omega$, which e.g. is not the case in open systems;
ii) it requires the map to be invertible;
iii) the proposal can only propose states that are forward or backward iterations of $\vx$. Consequently, for the random walk to be ergodic in the phase-space, the map itself must be ergodic.
iv) this proposal diffuses without drift on a trajectory passing through $\vx$, by shifting the starting point forward or backward. Thus, it will always sample fewer states than a time average of a trajectory starting at $\vx$.
On the other hand, the main advantage of this proposal is that it performs non-local jumps in the phase-space.
That is, it allows to jump from a region of the phase-space to another region while still maintaining $\vx'$ correlated with $\vx$.
As shown below, in combination with other proposal, this proposal is useful to reduce correlations stemmed from local jumps.

\subsubsection{Neighbourhood proposal}

\newcommand*\deltax{\delta_x(\vx)}
\newcommand*\vdelta{\boldsymbol{\delta}}
\newcommand*\vdeltax{\vdelta(\vx)}
\newcommand*\vddeltax{\hat{\vdelta}(\vx)}
\newcommand*\vddelta{\hat{\vdelta}}

Another strategy to construct a proposal on which on average the states are correlated by $\tstar(\vx)$ is to perturb $\vx$ by a finite amount $\vdelta$, $\vx' = \vx + \vdelta$, characterised by a direction $\vddelta$ and a norm $\delta$, $\vdelta \equiv \vddelta \delta$.
A common case is when the probability distribution is separated in two independent terms~\cite{RobertCasellaBook}:
\begin{equation}
P(\vdelta|\vx) = P(\hat{\vdelta} | \vx) P(\delta | \vx)
\end{equation}
and that $P(\hat{\vdelta} | \vx)$ is uniformly distributed in the $D$ directions and $P(\delta | \vx)$ has zero mean (i.e. an isotropic proposal).
Here we restrict the analysis to this situation, and we also assume that $P(\delta | \vx)$ is characterised by a well defined scale, e.g. it is an half-normal distribution\footnote{An exponential distribution would be equally acceptable and would not change the main conclusions.} with mean $\deltax$:
\begin{equation}
P(\delta | \vx) = \frac{\sqrt{2}}{\sqrt{\pi \sigma^2}} e^{- \frac{\pi \delta^2}{4\deltax^2}} \text{ for } \delta > 0 \ \ .
\label{eq:isotropic_proposal}
\end{equation}
This choice makes the ratio $g(\vx|\vx')/g(\vx'|\vx)$ in Eq.~\ref{eq:acceptance} to be
\begin{equation}
\frac{g(\vx|\vx')}{g(\vx'|\vx)} = \frac{\deltax}{\delta_x(\vx')} \exp \left[ -\frac{\pi|\vx' - \vx|^2}{4 \deltax^2} \left(1 - \frac{\deltax^2}{\delta_x(\vx')^2} \right) \right]
\label{eq:proposal_ratio}
\end{equation}
The main motivation for this choice is that the proposal distribution is described by a single function, $\deltax$, that quantifies the distance $\vx'-\vx$, $\Exp{|\vx' - \vx| |\vx} = \deltax$.

The goal is now to relate $\deltax$ with $\tstar(\vx)$.
Let us start to describe two important limits: in the limit $\deltax \rightarrow 0$, the states are the same and therefore $\lim_{\deltax \rightarrow 0} \tstar(\vx) = \tobs$.
In the limit $\deltax \rightarrow |\Gamma|$, the proposal is approximately equal to draw $\vx'$ uniformly from $\Gamma$, and $\vx'$ is independent of $\vx$ and $\tstar(\vx) = 0$.
To preserve a correlation of $\tstar(\vx)$, it is necessary that $\deltax$ is such that the two trajectories starting at $\vx$ and $\vx'$ are close together up a time $\tstar(\vx)$, see left panel of Fig.~\ref{fig:tstar}.
Because the system is chaotic, for small $\deltax$, two trajectories diverge exponentially in time according to Eq.~\ref{eq:divergence}, and, in particular, their maximal distance is given by Eq.~\ref{eq:exp_divergence}.
Therefore, to guarantee that two trajectories are distanced at most by $\Delta$ after a time $\tstar(\vx)$, $\deltax$ must be given by
\begin{equation}
\deltax = \Delta e^{-\lambda_{\tstar}(\vx)\tstar(\vx)}\ \ .
\label{eq:deltax}
\end{equation}
This equation relates the parameter of the proposal distribution, $\deltax$, with the average correlation $\tstar(\vx)$ of the two states $\vx$ and $\vx'$.

The neighbourhood proposal derived above is closely related to a proposal described in Ref.~\cite{Grunwald2008} as ``precision shooting''.
Precision shooting constructs a trajectory $\{\vx_i'\}$ with $\vx_0' \equiv \vx' = \vx + \delta \hat{\vdelta}$ (where $\delta$ is a free parameter) that, within the numerical precision of a computer, is indistinguishable (in the system considered) from a trajectory starting at $\vx'$ with $\delta$ small.
The trajectory $\{\vx_i'\}$ shadows a true trajectory starting at $\vx'$, in the same spirit as the algorithm used in Ref.~\cite{Grebogi1990} to construct a pseudo-trajectory.
Thus, precision shooting can be interpreted as the neighbourhood proposal, Eq.~\ref{eq:deltax}, with $\tstar$ a free parameter (related to $\delta$ via Eq.~\ref{eq:deltax}), that assumes shadowing theorem to simplify the construction $\vx'$.
Ref.~\cite{Grunwald2008} discusses how the acceptance rate depends on $\delta$, suggesting that the acceptance rate increases with decreasing $\delta$ (Fig.~9 of the ref.).
In light of the discussion in section~\ref{sec:correlation}, this result is interpreted as follows: as $\delta$ decreases, $\vx'$ becomes more correlated with $\vx$ (since $\tstar$ is related with $\delta$ by Eq.~\ref{eq:deltax}), and therefore the acceptance is expected to increase, as indicated in Fig. 9 of Ref.~\cite{Grunwald2008}.
This discussion is unfortunately insufficient to us because it does not allow to derive $\delta$ (or $\tstar$) that fulfils the condition in Eq.~\ref{eq:energy_condition}.
The crucial advantage of Eq.~\ref{eq:deltax} is that it allows to relate $\Exp{E(\vx') - E(\vx)|\vx}$ (in Eq.~\ref{eq:energy_condition}) with $\vx' - \vx$ (in Eq.~\ref{eq:deltax}) through $\tstar$ (in Eq.~\ref{eq:tstar}).
This is the goal of the next section.

\subsection{Guarantee local proposals}

Now that we derived proposal distributions that enforce an average correlation $\tstar(\vx)$, the next (and final) step is to obtain a relationship between $\tstar(\vx)$ and $\Exp{E(\vx') - E(\vx)|\vx}$.
Because the computation of $\tstar(\vx)$ depends on the particular observable $E$, a different derivation is presented for two observables, $t_e$ and $\lambda_{\tobs}$.
Given the limitations of the shift proposal described above, the argumentation below is
limited to neighbourhood proposals (an equivalent argumentation can be made to the shift proposal).

\subsubsection{FTLE in closed systems}
\label{sec:tstar_ftle}

As introduced in section~\ref{sec:ftle}, the observable in this case is given by $E(\vx) = \tobs\lambda_{\tobs}(\vx)$, where $\tobs$ is the finite time.
The aim in this case is thus to write $\Exp{\tobs\lambda_{\tobs}(\vx') - \tobs\lambda_{\tobs}(\vx) |\vx}$ as a function of $\tstar(\vx)$.
The finite-time Lyapunov exponent considered in
Eq.~\ref{eq:finite_time_directional_lambda_sum} is a sum of $\tobs$ terms\footnote{This equation is strictly valid for 1D, but the idea is that we split the FTLE between a component up to $\tstar$ and the remaining component.} and thus $\tobs \lambda_{\tobs}(\vx)$ can be written as the sum of the FTLE up to time $\tstar$ and the FTLE from $\tstar$ up to $\tobs$,
\begin{equation}
\tobs \lambda_{\tobs}(\vx) = \tstar \lambda_{\tstar}(\vx) + (\tobs - \tstar) \lambda_{\tobs - \tstar}(\vx_{\tstar}) \ \ ,
\label{eq:expanded_lyapunov_x}
\end{equation}
where $\vx_{\tstar} \equiv \vF^{\tstar}(\vx)$.
Likewise for the the trajectory starting from $\vx'$,
\begin{equation}
t_o\lambda_{t_o}(\vx') = \tstar \lambda_{\tstar}(\vx') + (t_o - \tstar) \lambda_{t_o-\tstar}(\vx_{\tstar}') \ \ .
\label{eq:expanded_lyapunov_xprime}
\end{equation}
Because $\vx'$ is proposed according to Eq.~\ref{eq:deltax}, by construction, the first $\tstar$ states of the trajectory starting at $\vx'$ are close (within $\Delta$) to the states of the trajectory starting at $\vx$ up to $\tstar(\vx)$. 
Therefore, we can approximate that the respective Lyapunovs up to time $\tstar$ are equal,
\begin{equation}
\Exp{\lambda_{\tstar}(\vx')|\vx} \approx \lambda_{\tstar}(\vx) \ \ .
\label{eq:approximation_equal_lyapunovs}
\end{equation}
Subtracting Eq.~\ref{eq:expanded_lyapunov_x} from Eq.~\ref{eq:expanded_lyapunov_xprime} and using Eq.~\ref{eq:approximation_equal_lyapunovs} gives
\begin{multline}
\Exp{\tobs\lambda_{\tobs}(\vx') - \tobs\lambda_{\tobs}(\vx) | \vx} =  \\
(\tobs - \tstar) \left( \Exp{\lambda_{\tobs - \tstar}(\vx_{\tstar}') | \vx} - 
\lambda_{\tobs - \tstar}(\vx_{\tstar})
\right) \ \ .
\label{eq:tstar_lyapunov_step}
\end{multline}
The left side of this equation is the same as in Eq.~\ref{eq:energy_condition} and thus
the aim now is to write the right side as a function of properties of the system.
Let us focus on the calculation of $\Exp{\lambda_{\tobs - \tstar}(\vx_{\tstar}') | \vx}$ first.
By construction, $\vx'$ is generated such that $|\vx_{\tstar}' - \vx_{\tstar}| \approx \Delta$.
Because the system is chaotic, one can approximate that $\vx_{\tstar}'$ is sufficiently separated from $\vx_{\tstar}$ such that $\lambda_{\tobs - \tstar}(\vx_{\tstar}')$ is independent of $\vx$.
Under this approximation, $\vx_{\tstar}'$ is essentially a random state from the phase-space, and thus $\lambda_{\tobs - \tstar}(\vx_{\tstar}')$ will be a drawn from the distribution of FTLE at time $\tobs - \tstar$.
Denoting the mean of this distribution by $\lambda_{L,\tobs - \tstar}$, we get
\begin{equation}
\Exp{\lambda_{\tobs-\tstar}(\vx_{\tstar}')|\vx} = \lambda_{L,\tobs - \tstar} \ \ .
\label{eq:approximation_independent_lyapunovs}
\end{equation}
Replacing Eq.~\ref{eq:approximation_independent_lyapunovs} in Eq.~\ref{eq:tstar_lyapunov_step} gives
\begin{multline}
\Exp{\tobs\lambda_{\tobs}(\vx') - \tobs\lambda_{\tobs}(\vx) | \vx} = \\ (\tobs - \tstar(\vx)) \left(\lambda_{L,\tobs - \tstar} - \lambda_{\tobs-\tstar}(\vx_{\tstar}) \right) \ \ .
\label{eq:tstar_lyapunov_step1}
\end{multline}
This equation relates the expected change in the observable with properties of the system ($\lambda_{L,\tobs - \tstar}$), of the trajectory $\vx$, $\lambda_{\tobs-\tstar}(\vx_{\tstar})$, and $\tstar(\vx)$ and it can thus be used in the energy condition we obtained earlier, Eq.~\ref{eq:energy_condition}.
Replacing the left side of Eq.~\ref{eq:energy_condition} by the expectation in Eq.~\ref{eq:tstar_lyapunov_step1} and solving to $\tstar(\vx)$ gives 
\begin{equation}
\tstar(\vx) = \tobs - \frac{a - 1}{d\log \p(\lambda_t(\vx))/d\lambda_t}\frac{1}{\lambda_{L,\tobs - \tstar} - \lambda_{\tobs-\tstar}(\vx_{\tstar})}\ \ .
\label{eq:general_tstar_lyapunov}
\end{equation}
We can further simplify this relation with two approximations: a) for large $\tobs - \tstar$, the mean FTLE at time $\tobs - \tstar$, $\lambda_{L,\tobs - \tstar}$, is approximately the Lyapunov exponent of the system, $\lambda_L$, 
\begin{equation}
\lambda_{L,\tobs - \tstar} \approx \lambda_L \ \ .
\label{eq:approximation_mean_lambda}
\end{equation}
b) because trajectories of chaotic systems are short-correlated in time, the FTLE of the trajectory $\vx$ up to $\tstar$ will be approximately equal to the FTLE of the trajectory starting at $\vx_{\tstar}$. Thus,
\begin{equation}
\lambda_{\tobs-\tstar}(\vx_{\tstar}) \approx \lambda_{\tobs}(\vx) \ \ .
\label{eq:approximation_equal_lambdas}
\end{equation}
Using Eq.~\ref{eq:approximation_mean_lambda} and Eq.~\ref{eq:approximation_equal_lambdas}, $\tstar(\vx)$ in Eq.~\ref{eq:general_tstar_lyapunov} can be written as
\begin{equation}
\tstar(\vx) = \tobs - \frac{a - 1}{d\log \p(\lambda_t(\vx))/d\lambda_t}\frac{1}{\lambda_L - \lambda_{\tobs}(\vx)}\ \ .
\end{equation}
To enforce that $\tstar \in [0, \tobs]$, we use
\begin{equation}
\tstar(\vx) = \max \left\{0, \tobs - \left | \frac{a - 1}{d\log \p(\lambda_t(\vx))/d\lambda_t}\frac{1}{\lambda_L - \lambda_{\tobs}(\vx)} \right | \right\} \ \ ,
\label{eq:tstar_lyapunov}
\end{equation}
which is the main result of this section.
This equation provides an expression to $\tstar(\vx)$ that can be inserted in the
parameter of the proposal distribution, Eq.~\ref{eq:deltax}, that under the approximations
used above achieves a constant acceptance rate\footnote{
In Ref.~\cite{Leitao2014} we used $\tstar = \tobs - 1$, different from Eq.~\ref{eq:tstar_lyapunov}.
The derivation of Eq.~\ref{eq:tstar_lyapunov} uses stronger approximations than the derivation of $\tstar = \tobs - 1$.
See argumentation after Eq.~(14) of Ref.~\cite{Leitao2014}.
}.

The derivation of Eq.~\ref{eq:tstar_lyapunov} can be generalised to other observables
which, as $\lambda_{\tobs}$, can be written as an average over the trajectory: consider
\begin{equation}
e_{\tobs}(\vx) \equiv \frac{1}{\tobs}\sum_{i=1}^{\tobs} f(\vx_i) = E_{\tobs}(\vx)/\tobs
\end{equation}
where $f(\vx_i)$ is an arbitrary function of the phase-space (the logarithm of the derivative of the map corresponds to $e_{\tobs}(\vx)=\lambda_{\tobs}(\vx)$, $E_{\tobs}(\vx)=\lambda_{\tobs}(\vx) \tobs$).
Replacing this quantity in the derivation of Eq.~\ref{eq:tstar_lyapunov} mutatis mutandis and without using the approximation in Eq.~\ref{eq:approximation_equal_lambdas}, one obtains
\begin{equation}
\tstar(\vx) = \max \left\{0, \tobs - \left | \frac{a - 1}{d\log \p(E_{\tobs}(\vx))/dE}\frac{1}{E_L - E_{\tobs-\tstar}(\vx)} \right | \right\} \ \ ,
\label{eq:tstar_energy}
\end{equation}
where $E_L$ is approximately the mean of the distribution $P(E_{\tobs})$ (using the approximation in Eq.~\ref{eq:approximation_independent_lyapunovs}).
This generalizes Eq.~\ref{eq:tstar_lyapunov} for an arbitrary average over trajectories of size $\tobs$, $e_{\tobs}(\vx)$, and it should be useful to sample rare states in respect to observables correspondent to expected values over trajectories.

\subsubsection{Escape time in strongly chaotic open systems}
\label{sec:derivation_te}

As introduced before, in strongly chaotic open systems $E(\vx) = t_e(\vx)$ and $P(t_e) \sim \exp(-\kappa t_e)$.
The aim here is to compute $\tstar(\vx)$ that fulfils Eq.~\ref{eq:energy_condition} by taking into account that $\vx$ is given and $\vx' = \vx + \vddelta \deltax$ with $\deltax$ given by Eq.~\ref{eq:deltax}.
In this case the trajectory's length is not a constant but instead it is given by the escape time $t_e(\vx)$.

A trajectory starting at $\vx'$ proposed according to Eq.~\ref{eq:deltax} fulfils $|\vx_{\tstar}' - \vx_{\tstar}| \leq \Delta$.
Therefore, up to $\tstar$, the two trajectories are indistinguishable.
Under this assumption, and because from the definition of $\tstar$ in Eq.~\ref{eq:tstar}, $t_e(\vx') = \tstar(\vx) + t_e(\vx_{\tstar}')$ and therefore 
\begin{equation}
\Exp{t_e(\vx')|\vx} = \tstar(\vx) + \Exp{t_e(\vx_{\tstar}')|\vx} \ \ ,
\label{eq:tstar_hyperbolic_step1}
\end{equation}
where $t_e(\vx_{\tstar}')$ is the escape time of the state $\vx'$ iterated $\tstar$ times
and  $\Exp{t_e(\vx_{\tstar}')|\vx} = \int_\Omega g(\vx'|\vx) t_e(\vx_{\tstar}') {\bf d}\vx'$
is the expected  $t_e(\vx_{\tstar}')$ over a neighbourhood of $\vx$ of size
$\delta_x(\vx)$ (given by Eq.~\ref{eq:deltax})). 
\begin{figure}[!t]
\centering
\includegraphics[width=\columnwidth]{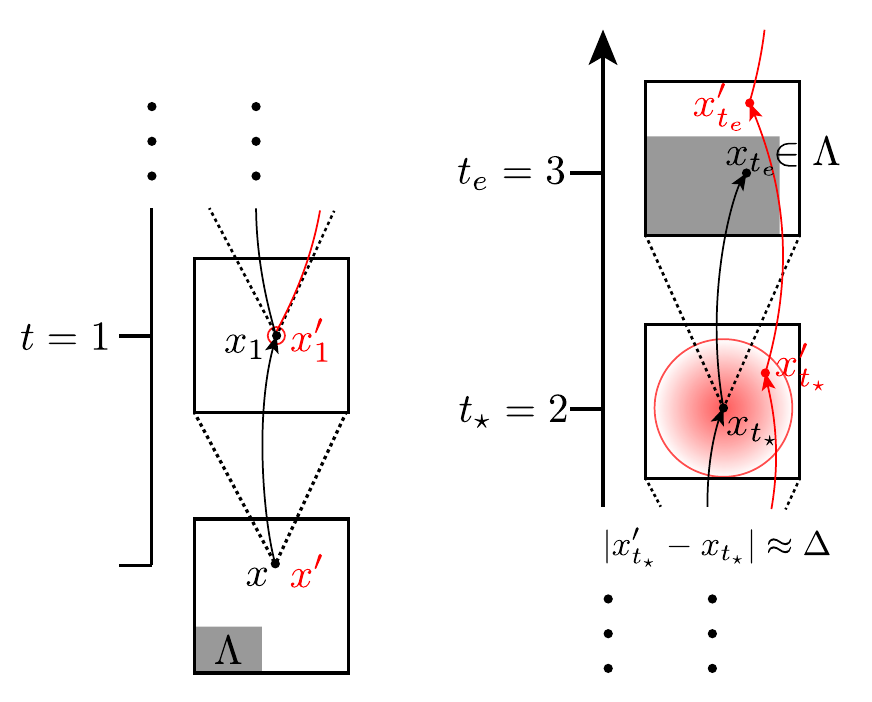}
\caption{
(Color online) Illustration of the core idea of our proposal for the case of the escape time:
a trajectory $\vx'$ starting close to $\vx$ and that remains similar to $\vx$ up to a time $\tstar=2$ will likely leave at a time $t_e(\vx') > t_e(\vx) = 3$.
The black (red) lines represent the iteration of the state $\vx$ ($\vx'$) until it leaves at time $t_e(\vx) = 3$ ($t_e(\vx') > 3$) by entering the shaded exit region~$\Lambda$.
The squares represent zooms of $\exp(\lambda_t(\vx) t)$ to the phase-space and thus correlated trajectories up to $\tstar$ correspond to select a specific zoom.
Our proposal is constructed in such a way that 
1) the distance between $\vx$ and $\vx'$  at time $\tstar$ is on average $\Delta$  and
2) $\tstar$ is chosen to be neither too large nor too small.
}
\label{fig:escape_time_close_to_hole}
\end{figure}
The idea behind this equation is represented in Fig.~\ref{fig:escape_time_close_to_hole}.
The proposal density $g(\vx'|\vx)$ at $\vx$ is such that, at $\tstar$, the two trajectories distance themselves on average by $\Delta$.
After $\tstar$, the trajectory starting at $\vx'_{\tstar}$ will approximately be independent of $\vx$ and thus $t_e(\vx') = \tstar + t_e(\vx'_{\tstar})$.
Moreover, proposing according to Eq.~\ref{eq:deltax} is equivalent to ``zoom'' the landscape of $t_e(\vx)$ around $\vx$ with a scale correspondent to $\tstar$'th iteration
of the construction of the landscape. Given the self-similarity of the landscape (see e.g. Fig.~\ref{fig:open_tent_landscape}), under this zoom, the landscape of $t_e(\vx_{\tstar})$ is equal to the landscape of $t_e(\vx)$ and therefore $\Exp{t_e(\vx_{\tstar}')|\vx}$ should be a constant independent of $t_e$.
Moreover, because $\vx_{\tstar}'$ is approximately independent of $\vx$, 
$\Exp{t_e(\vx_{\tstar}')|\vx}$ is the just average escape time of an independent state $\vx'_{\tstar}$, which is the average of $P(t_e)$ and is given approximately  by $1/\kappa$.
Thus, $\Exp{t_e(\vx_{\tstar}')|\vx}$ is the average escape time of an independent state $\vx'_{\tstar}$, which is the average of $P(t_e)$ and is given by $1/\kappa$:
\begin{equation}
\Exp{t_e(\vx_{\tstar}')|\vx} = 1/\kappa \ \ .
\label{eq:tstar_hyperbolic_step3}
\end{equation}
It is this result that incorporates the self-similarity of the escape time function: the value of $\tstar$ chooses the particular zoom of the landscape (Fig.~\ref{fig:escape_time_close_to_hole}), and this equation assumes that, as long as the zoom is proportional to $\exp(\lambda \tstar)$, the average escape time of the phase-space of the zoomed region is still $1/\kappa$.
Replacing Eq.~\ref{eq:tstar_hyperbolic_step3} in Eq.~\ref{eq:tstar_hyperbolic_step1} gives
\begin{equation}
\Exp{t_e(\vx')|\vx} = \tstar(\vx) + 1/\kappa \ \ ,
\label{eq:assumption_tstar_kappa}
\end{equation}
and subtracting $t_e(\vx)$ on both sides of Eq.~\ref{eq:assumption_tstar_kappa} gives
\begin{equation}
\mathbb{E}\left[ t_e(\vx') - t_e(\vx) | \vx \right] = \tstar(\vx) + \frac{1}{\kappa} - t_e(\vx) \ \ .
\label{eq:tstar_energy_relation_hyperbolic}
\end{equation}
The left side of this equation is the left side of the condition of constant acceptance rate, Eq.~\ref{eq:energy_condition}.
Equating both left sides and solving for $\tstar(\vx)$ gives
\begin{equation}
\tstar(\vx) = t_e(\vx) - \frac{1}{\kappa} - \frac{a - 1}{d\log \p(t_e)/dt_e} \ \ ,
\label{eq:tstar_hyperbolic}
\end{equation}
which is analogous to Eq. \ref{eq:tstar_lyapunov} and is the central result of this section.
Together with Eq.~\ref{eq:deltax}, it is the proposal distribution we had the goal of constructing.

\subsection{Summary: how to propose}
\label{sec:formulas_summary}

The argument in this section can be summarised as follows:
\begin{itemize}
\item Metropolis-Hastings requires a proposal that guarantees a specific variation of $E$, which we approximate by Eq.~\ref{eq:energy_condition}.
\item the proposal with a scale given by Eq.~\ref{eq:deltax} guarantees that on average the trajectory starting at $\vx'$ stays close to the trajectory starting at $\vx$ up to $\tstar(\vx)$.
\item  the variation of $E$ is related to $\tstar(\vx)$ via Eqs.~\ref{eq:tstar_lyapunov_step1} and~\ref{eq:tstar_energy_relation_hyperbolic}.
\end{itemize}
These three steps led to analytical formulas for $\tstar(\vx)$, Eqs.~\ref{eq:tstar_hyperbolic} and~\ref{eq:tstar_lyapunov}, that make the proposal satisfy Eq.~\ref{eq:energy_condition}.
Together with Eq.~\ref{eq:deltax}, they define proposal distributions required for an average constant acceptance.

There are 3 points that deserve to be noted:
the first point is that the formulas we obtained for $\tstar(\vx)$ are also valid for the problem of \emph{finding rare states} discussed e.g. in Refs.~\cite{Sweet2001,Tailleur2007,Kitajima2011,Iba2014,Bollt2005}.
Specifically, these formulas were derived to guarantee Eq.~\ref{eq:energy_condition}, which dictates how different $\Exp{E(\vx')|\vx}$ has to be from $E(\vx)$ to guarantee a constant acceptance.
This condition is stronger than the condition required for an algorithm to find minima or maxima of $E$, which requires only proposing states $\vx'$ such that $\Exp{E(\vx')|\vx} > E(\vx)$ (or vice-versa for minimising $E$).
This is independent of the particular minimisation algorithm (e.g. stimulated annealing, step descent, stagger and dagger in open systems) because it only discusses which new state $\vx'$ should be \emph{tried}, given the current state $\vx$.
The second point is that Eqs.~\ref{eq:tstar_hyperbolic} and~\ref{eq:tstar_lyapunov} reduce the proposal to a uniform distribution when the sampling distribution is the uniform distribution: $d\log \p(E)/dE(E) \rightarrow 0$ implies $\delta_x(\vx) \rightarrow \infty$.
The third point is that different proposals in the literature, precision shooting~\cite{Grunwald2008,Leitao2014}, exponential proposal distribution~\cite{Sweet2001,Kitajima2011}, and the one in Ref.~\cite{Leitao2013}, can be obtained from the proposals derived in this section through different approximations, as shown in Appendix~\ref{app.simplifications}.

Overall, this section described a framework to add information of chaotic systems, in this case the self-similar properties of the landscape, the exponential divergence of trajectories, and the exponential decay of correlations, to construct an efficient Metropolis-Hastings algorithm to sample them. A simplified description of the algoirthm is given in in Appendix~\ref{app.algorithm} and an open-source implementation in Ref.~\cite{github}.
The next section is devoted to test the assumptions used here on each of the problems, escape time and FTLE, and confirm the practical usefulness of the framework.

\section{Numerical Tests}
\label{sec:applications}

The previous section concluded with a set of formulas -- the proposal in
Eq.~\ref{eq:deltax} combined with the formulas for $\tstar(\vx)$
Eq.~\ref{eq:tstar_lyapunov} or Eq.~\ref{eq:tstar_hyperbolic} -- for proposing states
$\vx'$ that are expected to guarantee the desired acceptance rate (bounded from $0$ and $1$). In this section we
test some of the approximations made in the derivation of $\tstar(\vx)$ (in a simple system) and we analyse the efficiency of the algorithm (confirming
the polynomial scaling) in the computation of the FTLE in closed system -- introduced in
Sec.~\ref{sec:ftle} -- and of the escape time in open systems -- introduced in
Sec.~\ref{sec:open_systems}.
The tests are performed in the skewed tent map and open skewed tent map, and the efficiency is tested in numerous maps (chain of couple H\'enon maps, Standard map, Logistic map). All these systems are introduced in detail in Appendix~\ref{app:maps}.

\subsection{Finite-time Lyapunov exponent}

The first approximation made in the derivation of Eq.~\ref{eq:tstar_lyapunov} is that when $\delta_x \equiv |\vx' - \vx|$ is drawn from a half-normal distribution with scale parameter $\deltax$ given by Eq.~\ref{eq:deltax}, $\vF^{\tstar}(\vx')$ is sufficiently close from $\vF^{\tstar}(\vx)$ such that $\Exp{\lambda_{\tstar}(\vx')|\vx} = \lambda_{\tstar}(\vx)$, Eq.~\ref{eq:approximation_equal_lyapunovs}, holds.
We test this numerically by randomly drawing $10^5$ states $\vx_i$ in the tent map with $a=3$ (see Appendix~\ref{app:maps}), and, for each, propose a state $\vx_i' = \vx_i + \vdeltax$ according to Eq.~\ref{eq:isotropic_proposal} with $\deltax$ given by our Eq.~\ref{eq:deltax}. 
From the $10^5$ pairs of states $(\vx_i,\vx'_i)$, we estimate the difference of the observables: $E' \equiv \Exp{t \lambda_t(\vx_i')|\vx}$ and $E \equiv t \lambda_t(\vx_i)$.
Our expectation is that for a fixed $\tstar = t$, $E' - E$ should be much smaller than $E$ (the relevant scale in Eq.~\ref{eq:tstar_lyapunov_step}).
We numerically obtain that within the $99\%$ quantile, $E' - E \approx 1$ independently of $E$ and $t$. This value is much smaller than $E \in [0.6t, 1.1t]$, specially since we are interested in large $t$.
This strongly supports the approximation we make in Eq.~\ref{eq:approximation_equal_lyapunovs}.

\begin{figure}[!t]
\centering
\includegraphics[width=0.8\columnwidth]{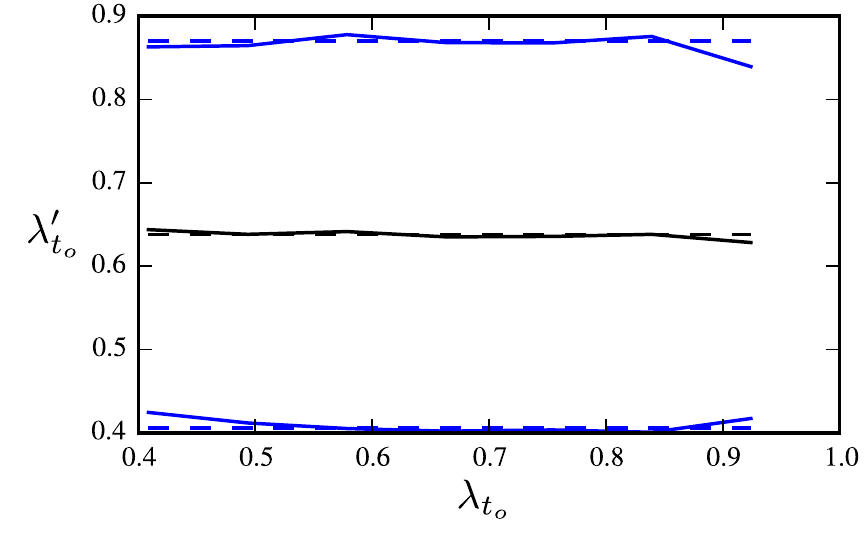}
\caption{
The finite-time Lyapunov exponent of the first half of the trajectory is independent from the one of the second half of the trajectory.
The graph was obtained by sampling $10^5$ random initial conditions $\vx_i$ and compute $\left( \lambda_t(\vx_i), \lambda_t(\vF^t(\vx_i)) \right)$ = (first half, second half) of a $2t=16$ steps trajectory.
The y axis represents the mean (full black) $\pm$ 2 standard deviations (full blue) of $\lambda_t(\vF^t(\vx_i))$ conditioned to a given $\lambda_t$.
The dashed lines represent the same mean and standard deviation, but over all points (without conditioning).
When $\lambda_t(\vF^t(\vx_i))$ is independent of $\lambda_t(\vx_i)$, the dashed and full lines are the same within fluctuations, as observed.
A Kolmogorov-Smirnov test comparing the un-conditioned and conditioned distributions gives a p-value higher than $0.001$ (hypothesis that they are independent is not rejected).
}
\label{fig:lambda_correlation_tent3}
\end{figure}

The second approximation made in the derivation of Eq.~\ref{eq:tstar_lyapunov} is that there is no dependence between $\lambda_{\tstar}(\vx)$ and $\lambda_{\tobs - \tstar}(\vF^{\tstar}(\vx))$, Eq.~\ref{eq:approximation_independent_lyapunovs}.
In other words, that the FTLE of the trajectory starting at $\vF^{\tstar}(\vx)$ and ending at $\tobs$ is indistinguishable from the one drawn from
the distribution of FTLE with finite-time $\tobs - \tstar$, $P(\lambda_{\tobs-\tstar})$.
This approximation was numerically tested by drawing points $\vx_i$, computing the pairs
$\lambda_t(\vx_i), \lambda_t(\vF^t(\vx_i))$ (i.e. $2\tstar = \tobs = 2t$), and testing
whether the conditional probability of $\lambda_t(\vF^t(\vx_i))$ equals the
(unconditional) probability of $\lambda_t(\vx_i)$.
The results in Fig.~\ref{fig:lambda_correlation_tent3} confirm the equality of these probabilities.

\begin{figure}[!t]
\centering
\includegraphics[width=0.8\columnwidth]{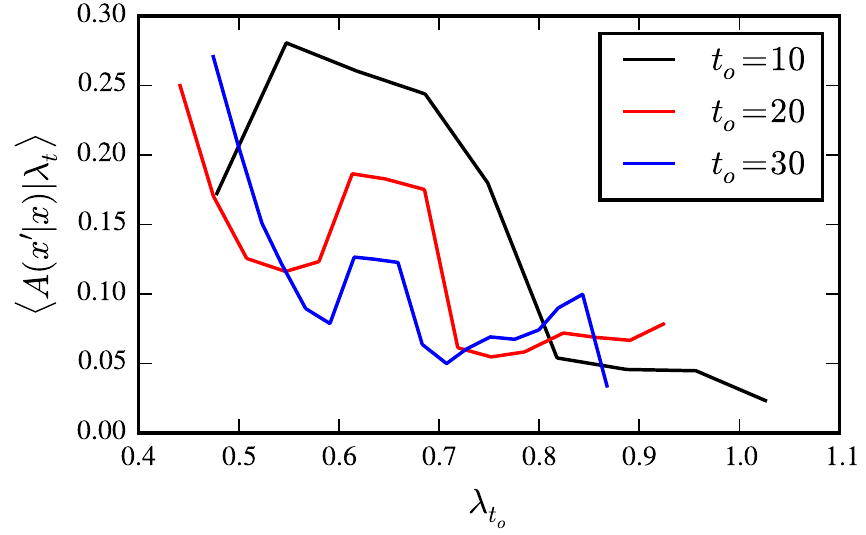}
\includegraphics[width=0.8\columnwidth]{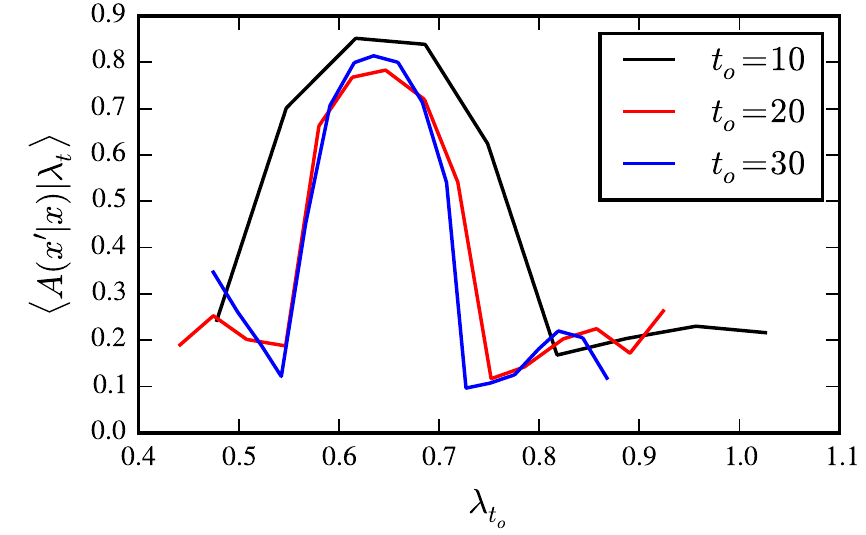}
\caption{
The $\tstar(\vx)$ given by Eq.~\ref{eq:tstar_lyapunov} guarantees a bounded acceptance ratio independently of $\tobs$.
The graph represents the average acceptance as a function of $\lambda_{\tobs}(\vx)$ obtained from uniformly sample of $10^6$ pairs of states $(\vx, \vx'(\vx))$ (see text for details) in the tent map with $a=3$, for different finite-times $\tobs$ and sampling distributions.
Top panel: $\p(\vx)$ is the canonical ensemble, Eq.~\ref{eq:canonical_ensemble}, with $\beta=1$.
Bottom panel: $\p(\vx)$ is the flat-histogram ensemble, Eq.~\ref{eq:flathistogram_ensemble}.
}
\label{fig:constant_acceptance_tent3}
\end{figure}

We finally test whether a proposal using $\tstar(\vx)$ given by
Eq.~\ref{eq:tstar_lyapunov} guarantees a constant acceptance rate, the original motivation
for our calculation. 
The test consisted in sampling $10^6$ states according to the following procedure: 1)
uniformly draw a state $\vx_i$ and compute $\lambda_i \equiv \lambda_{\tobs}(\vx_i)$; 2)
generate a state $\vx'_i$ according to the proposal distribution
Eq.~\ref{eq:isotropic_proposal} with $\deltax$ given by Eq.~\ref{eq:deltax} and $\tstar$
given by Eq.~\ref{eq:tstar_lyapunov} ($\delta_0 = 1$), and compute $\lambda_i' \equiv
\lambda_{\tobs}(\vx'_i)$; 3) store $G_i \equiv g(\vx|\vx')/g(\vx'|\vx)$ computed from
Eq.~\ref{eq:proposal_ratio}: $\delta_x$ is given by Eq.~\ref{eq:deltax}, and $|\vx' -
\vx|$ is given by storing $\delta_i = |\vx'_i-\vx_i|$. 
The ratio of the target distribution is given by $r_i \equiv \p(E')/\p(E) =
\exp(-\beta \tobs(\lambda_i' - \lambda_i))$ for the canonical ensemble and $r_i =
G(E)/G(E')$, where $G(E) = G(\tobs \lambda_i)$, is given by Eq.~\ref{eq:rho_lambda_tent_map} for the flat-histogram.
The numerically estimated acceptance ratio, $A(\lambda_t) \equiv
\left \langle \min(1, r_i \times G_i)\right \rangle$ is shown in Fig.~\ref{fig:constant_acceptance_tent3}. It is not independent of
$\lambda_{\tobs}$ -- the expected outcome based on the assumption of constant acceptance
used in our derivations -- but it is bounded from $0,1$ for increasing $N=\tobs$ --
the original requirement for an efficient simulation set in Sec. \ref{sec:derivation_ratio}.
In the canonical ensemble, there is linear dependency of $\Pi$ with $\lambda_{\tobs}$, and in the flat-histogram
ensemble, the ratio is 0.8 in the maximum of $P(\lambda_{\tobs})$, and decays to about $0.1$ on the tails.

\subsubsection{Efficiency of the flat-histogram ensemble}
\label{sec:efficiency_lambda}

\begin{figure}[!t]
\centering
\includegraphics[width=0.8\columnwidth]{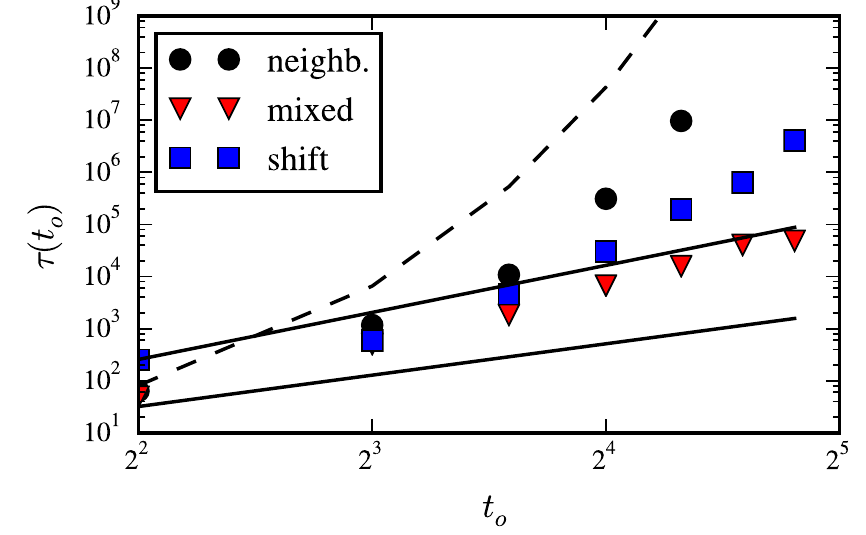}
\caption{Average round-trip of a flat-histogram in the tent map with a mixed proposal is
  polynomial, as opposed to uniform sampling, shift poposal, and neighbourhood proposal.
The simulation was made using a flat-histogram simulation on the tent map with $a=3$,
where $G(\lambda_t)$ is given from Eq.~\ref{eq:rho_lambda_tent_map}, and where the
round-trip time (efficiency) was defined as going from $\lambda_{\min}$ to $\lambda_{\max}$ and return.
The dashed black line represents $1/P(\lambda_{\max})$, the expected number of samples
required in uniform sampling; the bottom full line is proportional to $\tobs^2$, the upper
line is proportional to $\tobs^3$.
The shift proposal used  $\tshift = 1$ and  for the backward iteration in time it used one
(randomly chosen) of the two pre-images of the state. In the mixed proposal the shift and
neighbourhood proposals were chosen with probability $1/2$.
}
\label{fig:round-trip-lambda}
\end{figure}

The success in achieving a bounded acceptance independent of $N$, as the one confirmed in the previous section for
the tent map, does not guarantee the efficiency of the method (see
Sec.~\ref{sec:efficiency}).  Here we test the efficiency of flat-histogram computations of
$P(\lambda_t)$ in the tent map that use the neighbourhood proposal, the shift proposal,
and both (mixed proposal). The results shown in Fig.~\ref{fig:round-trip-lambda} suggest that only when
both proposals are used the efficiency scales polynomially with $N=\tobs$.

\begin{figure}[!t]
\centering
\includegraphics[width=\columnwidth]{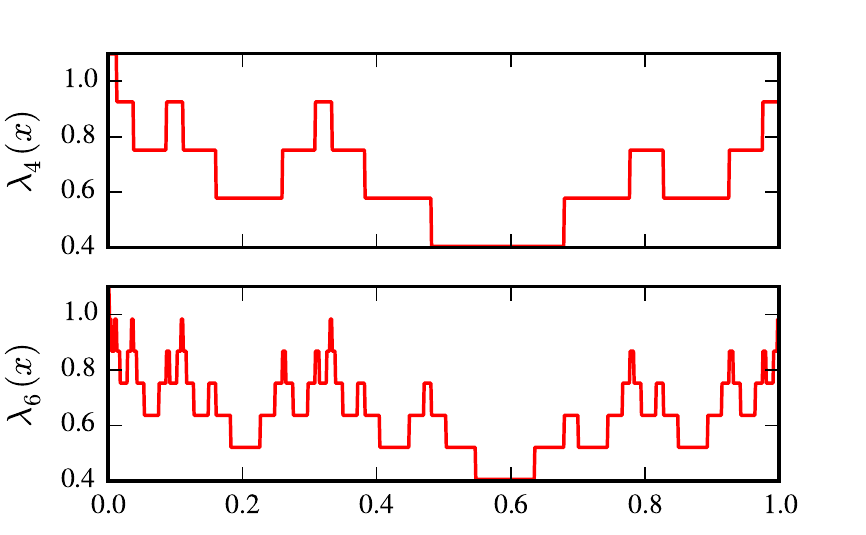}
\caption{
The finite-time Lyapunov exponent $\lambda_{\tobs}(x)$ in the tent map, Eq.~\ref{eq:lambda_tent_map}, already contains an increasing number of local minima and maxima with increasing $\tobs$, a crucial feature present in Fig.~\ref{fig:challenges_lyapunov}.
In both cases, the maximum $\lambda_{\tobs}(x)$ is $\lambda_{\max} \approx 1.09$, and the minimum is $\lambda_{\min} \approx 0.4$.
}
\label{fig:landscape_tent}
\end{figure}

This result can be understood looking at the landscape of $\lambda_t$, as illustrated in
Figure~\ref{fig:landscape_tent}. Imagine a flat-histogram simulation on this system, for $\tobs=4$ (black curve in the figure), and analyse what happens to it in terms of a round-trip.
Lets suppose that the simulation was recently at the minimum $\lambda_t$ (0.41) and that the next round-trip is made by going to the maximum $\lambda_t$ (1.09) and return back.
Lets further suppose that the simulation eventually got to a state with $\lambda_t \approx 0.92$.
Because $\p(\vx)=\p(\lambda_t(\vx))$, every state at that $\lambda_t$ is equiprobable.
Therefore, the state can be at any plateau (of the 4, see fig.), proportionally to their plateau-size.
However, not every plateau contains, on its neighbourhood, a neighbour plateau with higher $\lambda_t$, for example, the plateau around $0.3$.
Therefore, a local proposal would never be able to reach a higher plateau from a state on such a plateau.
First, it would need to go backward, reach the maximum of $P(\lambda_t)$ (around $0.69$), where the proposal proposes any other state, and then try to find another path towards a higher $\lambda_t$.
This would already be the problem if the simulation would be a canonical ensemble with a $\beta$ favouring higher $\lambda_t$'s, since it would require decreasing $\lambda_t$ by an amount $\Delta \lambda$, and this happens with a probability that decays exponentially with $\Delta \lambda$.
This is solved by using a flat-histogram ensemble.
However, the crucial challenge here is that as $t$ increases, the number of local maxima also increases, but the number of
maxima connected with the global maximum is constant: the red curve, with $t=6$, contains now $11$ plateaus for $\lambda_t \approx 0.87$, but only $1$ is locally connected to the maximum $\lambda_t$, the one around 0.
This means that, as $t$ increases, it becomes more difficult to perform a round-trip: not only because the expected time to diffuse increases (see Eq.~\ref{eq:round_trip_scaling}), but also because there are more times where the simulation diffuses forward and backward until it reaches the global maxima.
The shift proposal alleviates this problem by allowing non-local proposals in the phase-space.
A shift proposes a state $\vx'$ on a non-neighbourhood of $\vx$, which improves the probability of reaching higher or lower $\lambda$'s, which explains why a combined proposal has such a low round-trip time in this system.

\begin{figure}[!t]
\centering
\includegraphics[width=\columnwidth]{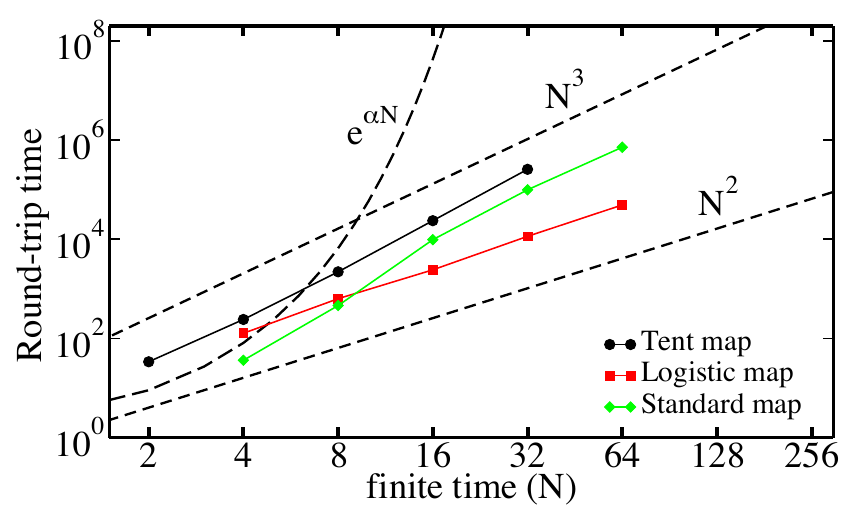}
\caption{
The number of samples required to sample a rare state, proportional to the round-trip time, scales polynomially with $N=\tobs$ in Metropolis-Hastings with the proposal distribution derived in Sec.~\ref{sec:efficiency}, as opposed to the exponential increase in uniform sampling.
Tent map: Eq.~(\ref{eq:tent_map}) with $a=3$; Logistic map: $F(x) = 4 x (1 - x)$; Standard map~\cite{OttBook}: $K=8$ and $\uni(\vx) = \text{const.}$ was used in every case.
The Wang-Landau algorithm was used to estimate the distribution prior to perform the flat-histogram and the distribution agrees with the analytical one when available~\cite{Prasad1999,BeckBook}.
The proposal distribution used was a mixed proposal composed by 50\% chance of being the neighbourhood proposal with $\tstar=\tobs - 1$ and $\Delta=1$, and 50\% chance of being the shift proposal with $\tshift = 1$.
Adapted from Ref.~\cite{Leitao2014}
}
\label{fig:lambda_efficiency}
\end{figure}

Finally, we confirm that a polynomial efficiency is obtained in the computation of $P(\lambda_{\tobs})$ more generally.
We performed flat-histogram simulations, using the Wang-Landau algorithm to estimate $P(\lambda_{\tobs})$ on different chaotic systems: the tent map, the logistic map, and the standard map (see appendix~\ref{app:maps} for details).
The results are shown in Fig.~\ref{fig:lambda_efficiency} and confirm the dramatic improvement and generality of using Metropolis-Hastings to sample rare states.
The simulations in Fig.~\ref{fig:lambda_efficiency} use $\tstar(\vx) = \tobs - 1$ instead of the one given by Eq.~\ref{eq:tstar_lyapunov}.
This is computationally always more expensive because the correlations due to the neighbourhood proposal are maximal (see the discussion after Eq.~\ref{eq:tstar_lyapunov}), but on the other hand this proposal is simpler because it does not require estimating $d\log P/dE$ and $\lambda_L$.

\subsection{Transient chaos}
\label{applications_te}

The derivation of Eq.~\ref{eq:tstar_hyperbolic} uses the assumption that when $\vx'$ is proposed with a scale given by Eq.~\ref{eq:deltax} with $\tstar = t_e$, $\Exp{ t_e(\vx') |\vx} = t_e(\vx)$, as per Eq.~\ref{eq:tstar_hyperbolic_step1}.
We tested assumption in a similar way we tested the approximation of Eq.~\ref{eq:approximation_equal_lyapunovs} for the FTLE, and consisted in uniformly drawing states $\vx_i$, compute their escape time $t_e \equiv t_e(\vx_i)$ and, for each, generate a state $\vx_i'=\vx_i+\vdeltax$ with a scale $\deltax$ given by Eq.~\ref{eq:deltax} with $\tstar(\vx) = t_e(\vx)$ and compute its escape time $t_e' \equiv t_e(\vx_i')$.
The assumption is valid when, on average, $t_e' - t_e \ll t_e$ for large $t_e$.
We did this experiment in the following systems: open tent map with $a=3$ and $b=5$, standard map with $K=6$, and coupled H\'enon maps with $D=2, 4, 6, 8$ (see appendix~\ref{app:maps}) for $\Delta = 1$.
We observed that in all 6 cases, the average $t_e' - t_e$ is smaller than 1 for all $t_e>2/\kappa$.
These observations show that the assumption of Eq.~\ref{eq:tstar_hyperbolic_step1} is valid for a broad class of chaotic systems.

A second assumption tested here is the self-similarity argument used in deriving Eq.~\ref{eq:tstar_hyperbolic_step3}.
The self-similarity assumption we use in Eq.~\ref{eq:tstar_hyperbolic_step3} is that the landscape is self-similar such that, irrespectively of the particular scale we choose (by decreasing $\tstar$), the properties of the zoomed phase-space remain the same.
In particular, we are interested in checking that $t_e' - t_e$ does not depend on $t_e$ when we choose a larger scale, i.e. when $\tstar$ is decreased by a constant.
We thus repeat the experiment above but we decrease $\tstar$ to $\tstar(\vx) = t_e(\vx) - T$ with $T=0, 1/\kappa, 5/\kappa$ to check that increasing $T$ decreases the average $t_e' - t_e$ without changing its independency with $t_e$.
Our numerical tests in the same systems as before show that $t_e' - t_e$ decreases with increasing $T$ and it remains independent of $t_e$, confirming our hypothesis.

\begin{figure}[!t]
\centering
\includegraphics[width=\columnwidth]{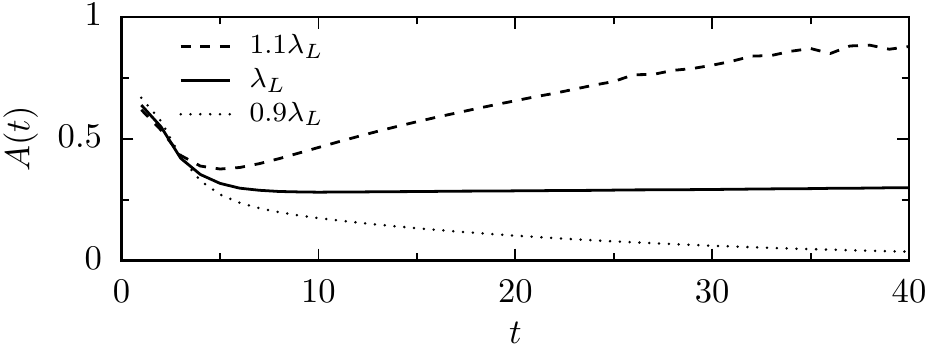}
\caption{
The acceptance rate of a Monte Carlo flat-histogram simulation is constant as a function of the escape time $t$ when the Lyapunov exponent of the system $\lambda_L$ is used.
The simulation was made on the open tent map with $a=3$ and $b=5$ with the exact $P(t_e)$ given by Eq.~\ref{eq:open_tent_distribution}.
Different curves represent using the proposal with $\deltax$ given by Eq.~\ref{eq:deltax_lyapunov} with three different exponents.
When the exponent is larger than $\lambda_L$, $\deltax$ effectively aims for a smaller $\tstar$ and therefore a larger distance $t_e'(\vx) - t_e(\vx)$, consequently decreasing the acceptance.
When the exponent is smaller than $\lambda_L$, $\deltax$ aims for a larger $\tstar$ than the one given by Eq.~\ref{eq:tstar_hyperbolic}, and therefore $t_e'(\vx) - t_e(\vx)$ decreases to 0 as $t_e(\vx) \rightarrow \infty$, and the acceptance converges to 1. Adapted from~\cite{Leitao2013}.
}
\label{fig:prl_fig2}
\end{figure}

The previous tests indicate that the proposal distribution should induce a constant acceptance rate when the Lyapunov exponent of the system, Eq.~\ref{eq:deltax_simplified}, is used.
To confirm that this is the case, a flat-histogram simulation with an isotropic proposal distribution with width $\deltax = \delta_x(t_e(\vx))$ given by Eq.~\ref{eq:deltax_simplified} with $\lambda_L(t_e) = \lambda_L$ was made.
The results, Fig~\ref{fig:prl_fig2}, reproduced from Ref.~\cite{Leitao2013}, confirm that proposing with $\lambda_L$ guarantees a constant acceptance, and that any other exponent in Eq.~\ref{eq:deltax_simplified} fails to achieve so.

The above tests confirm that the derivation made in Sec.~\ref{sec:derivation_te} holds for a paradigmatic strongly chaotic open system.
These tests also present a major advantage of using the approach in this paper: it allows to test the assumptions made on each step, something that other approaches, such as the ones in Refs.~\cite{Sweet2001,Kitajima2011,Laffargue2013a}, do not explicitly allow.

\begin{figure}[!t]
\centering
\includegraphics[width=\columnwidth]{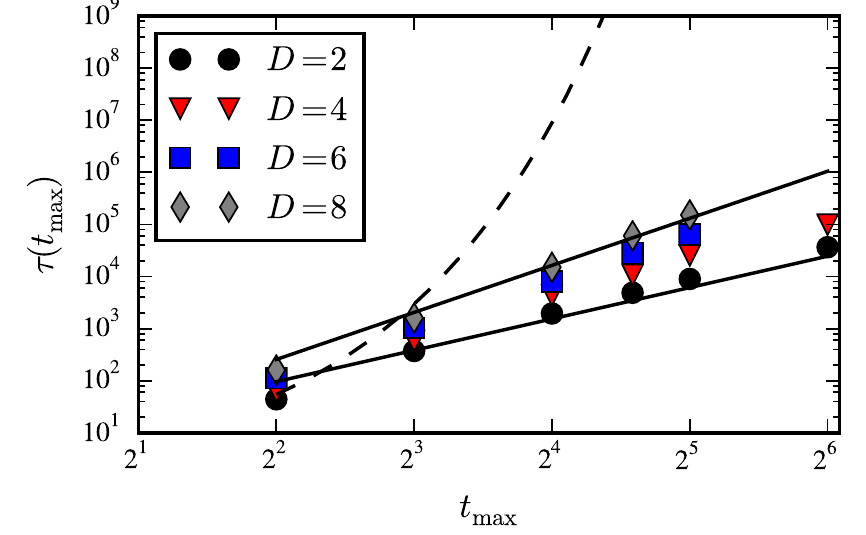}
\caption{
Polynomial scaling of the number of samples required to perform a round-trip ($1 \rightarrow t_{\max} \rightarrow 1$) as a function of $t_{\max}$ of the Metropolis-Hastings algorithm with the proposal derived in Sec.~\ref{sec:derivation_te}, as opposed to the exponential scaling in uniform sampling.
This plot represents the average round-trip time of a flat-histogram simulation with the proposal given in Sec.~\ref{sec:formulas_summary} and number of samples required to sample $t_{\max}$ in using uniform sampling (line) in the coupled H\'enon map, Eq.~\ref{eq:nhenon_map}, for different dimensions.
The two full lines represent $t_{\max}^2$ (lower) and $t_{\max}^3$ (upper), and the dashed line represents $1/P(t_e)$ for $D=4$ (e.g. from Fig.~\ref{fig:fractal_landscape}; $D>4$ have a even higher exponent).
The flat-histogram was obtained by first running a Wang-Landau algorithm for 10 refinement steps, each with 100 round-trips.
Each point represents the average round-trip time over 100 round-trips after the 10 refinement steps.
The proposal distribution used was the isotropic, Eq.~\ref{eq:isotropic_proposal}, with $\deltax$ given by Eq.~\ref{eq:deltax_lyapunov} with $\delta_0 = 10$.
}
\label{fig:prl_fig4}
\end{figure}

\begin{figure}
\centering
\includegraphics[width=\columnwidth]{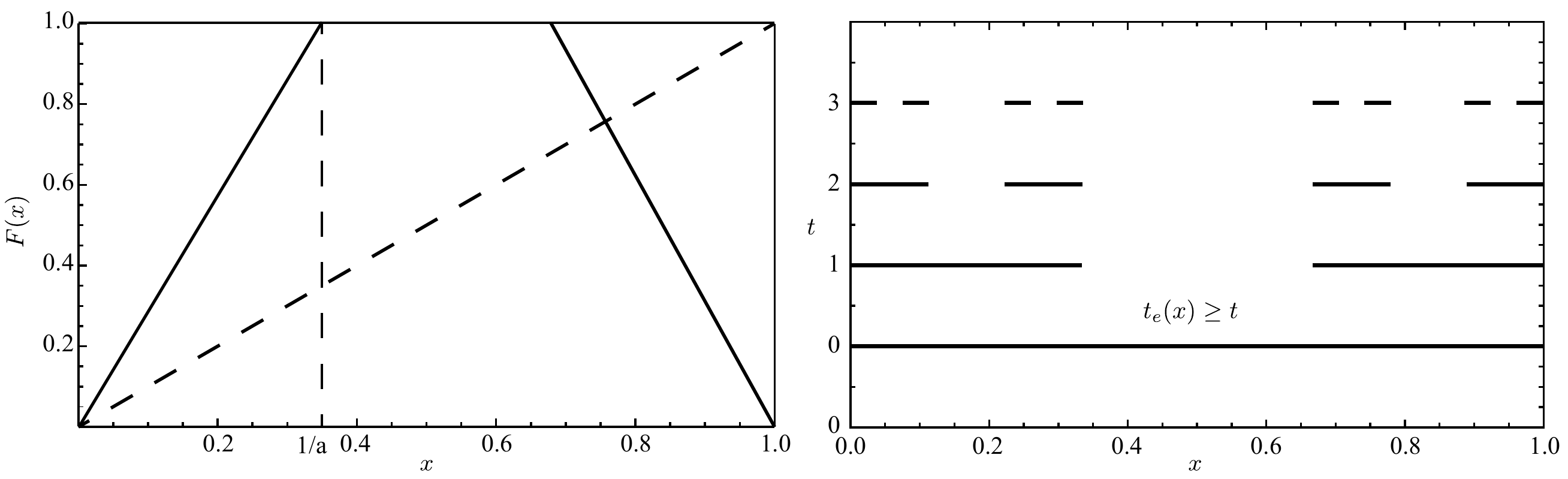}
\caption{
The open tent map, and its corresponding surviving set, equal to the construction of the cantor set.
(Left) The open tent map, where the escape correspond to states inside the interval in the middle, that maps to outside the unit interval.
(Right) An iteration of the open tent map with $a=b=3$ corresponds to remove the middle third of each of the plateaus of the surviving set at time $t$ and the set that survives this removal is the surviving set at $t+1$.
The third middle Cantor set is the surviving set at $t \rightarrow \infty$.
}
\label{fig:open_tent}
\end{figure}

\begin{figure}[!t]
\centering
\includegraphics[width=\columnwidth]{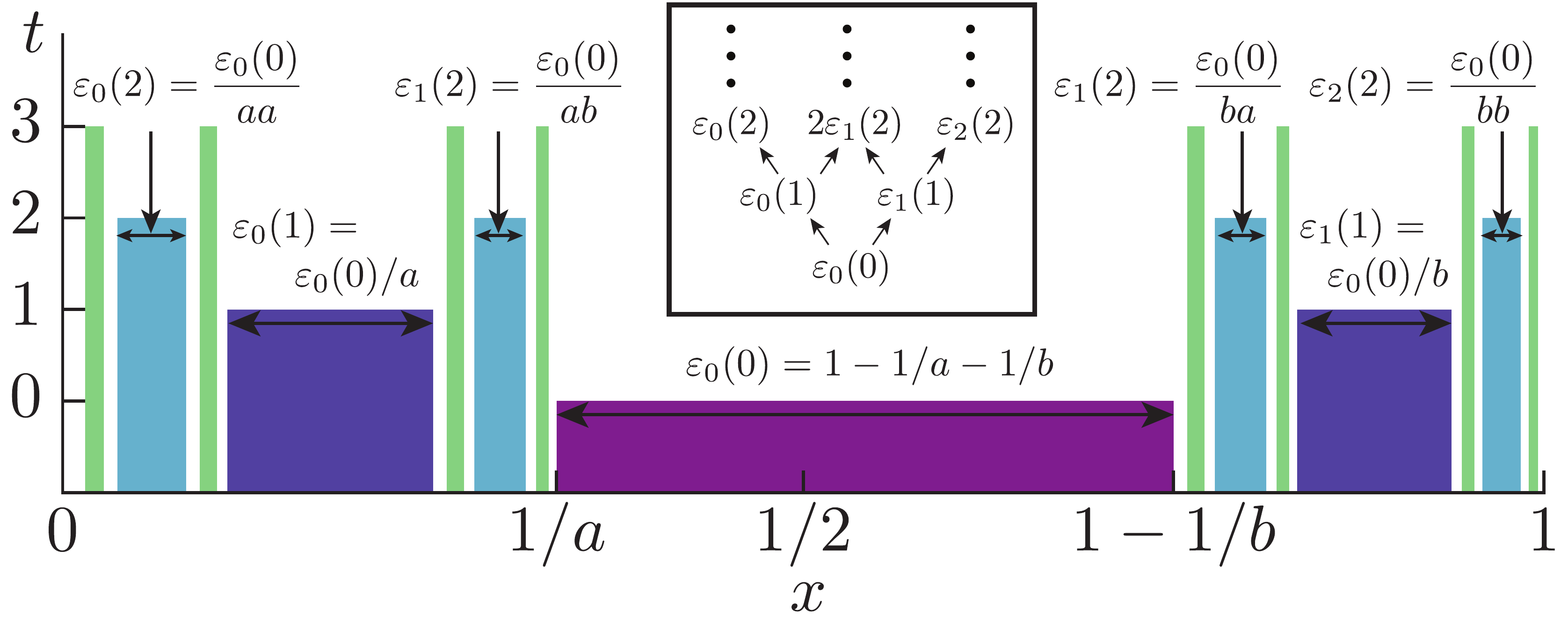}
\caption{
The escape time function, $t_e(x)$, of the tent map for a generic $a$ and $b$.
There are $2^t$ intervals, and the size of each plateau can be analytically computed from
the one at a previous time, and therefore be written analytically.
Specifically, a given interval has size $\varepsilon(x) = (1 - 1/a - 1/b)a^{-i(x)}b^{-t_e(x) + i(x)}$, where $i(x)$ is the number of times $0 < F^t(x) < 1/a$, for $t = 1,...,t_e(x)$.
The crucial observation is that $\lambda_{t_e(x)}(x)$ is proportional to $\log(\varepsilon(x))/t_e(x)$. Adapted from Ref.~\cite{Leitao2013}.
}
\label{fig:open_tent_landscape}
\end{figure}

The efficiency of the simulation is tested in Fig.~\ref{fig:prl_fig4} as a function of the maximal escape time considered, $t_{\max}$, $\tau(t_{\max})$ in the generic coupled H\'enon maps defined by Eq.~\ref{eq:nhenon_map}. It confirms the dramatic improvement of Metropolis-Hastings with the proposal derived in section~\ref{sec:open_systems} over uniform sampling: the scaling is polynomial using importance sampling, and exponential in uniform sampling.

The derivation in Sec.~\ref{sec:derivation_te}, the tests presented above, and results in Figure~\ref{fig:prl_fig4}, show the why and how importance sampling Metropolis-Hastings can efficiently sample long-living trajectories in strongly chaotic open systems.
The proposal distribution should be applicable to strongly chaotic open systems more generally, as the 
approximations made are expected to be valid in other strongly chaotic systems.


\section{Conclusions}
\label{sec:conclusions}

\subsection{Summary of results}
We have introduced a framework to sample rare trajectories in different classes of chaotic systems. It is based on the Metropolis-Hasting algorithm, a flexible and well-established Monte Carlo method suitable for the investigation of many numerical problems in chaotic dynamical systems (as shown in Sec.~\ref{sec:chaos}). Our main contribution is a procedure (see Sec.~\ref{sec:framework}) to construct the proposal step of the Metropolis-Hasting algorithm (which proposes a new state $\vx'$ given the current state $\vx$) that ensures the efficiency of the sampling. The main arguments in the construction of this procedure are:

\begin{itemize}

\item[(i)] set (in Sec.~\ref{sec:derivation_ratio}) as an heuristic goal to have a bounded (or constant) acceptance rate~(\ref{eq:acceptance}).
This generalizes the traditional heuristic~\cite{Lopes.phd2006} used in Metropolis-Hastings, $E(\vx') - E(\vx) \sim 1$ and can therefore be used more generally in Metropolis-Hastings simulations.\footnote{
For example, using this heuristics, in a Metropolis-Hastings flat-histogram in the Ising model (in this case $\vx$ corresponds to a list of all spins, $E(\vx)$ corresponds to the energy of the configuration), when the configuration of the system is close to the maximum of the density of states (maximal energy, no magnetization), there is no need to flip just one spin at the time: one can flip all spins at once because, there, correlating such states brings no advantage to increase the acceptance rate (it is going to be accepted anyway), but it increases $r$.}

\item[(ii)] introduce (in Sec.~\ref{sec:tstar}) an auxiliary quantity, the correlation time $\tstar$,  that quantifies the similarity between any two states. It is motivated by the notion that the observables considered in section~\ref{sec:chaos_summary} are computed over trajectories, and $\tstar(\vx,\vx')$ quantifies the time in which trajectories remain close to each other. We then showed how two proposal distributions, shift and neighborhood, can be used to control the average of $\tstar(\vx,\vx')$ over $\vx'$, $\tstar(\vx)$.

\item[(iii)] derive (in Sec. \ref{sec:shift_proposal}) an expression for the values of $\tstar(\vx)$ -- Eq.~\ref{eq:tstar_lyapunov} and \ref{eq:tstar_hyperbolic} -- which should be used in order to guarantees a constant acceptance. This is done for two observables of interest in chaotic systems (see Sec.~\ref{sec:stability}) -- the escape time $E(\vx) = t_e(\vx)$ and the finite-time Lyapunov exponent (FTLE) $E(\vx) = \lambda_{\tobs}(\vx)$ -- and two target distributions of the Metropolis Hasting method (see Sec.~\ref{sec:importance_sampling}) -- canonical and flat-histogram. These results are summarized in Tab.~\ref{tab:summary}.

\end{itemize}

A successful application of our framework leads to an algorithm in which the number of samples to obtain an independent rare sample scales polynomially with the difficulty of the problem, as opposed to the exponential increase observed in traditional uniform sampling (see Fig.~\ref{fig:lambda_efficiency} and \ref{fig:prl_fig4}).

\begin{table}
\centering
\includegraphics[width=\columnwidth]{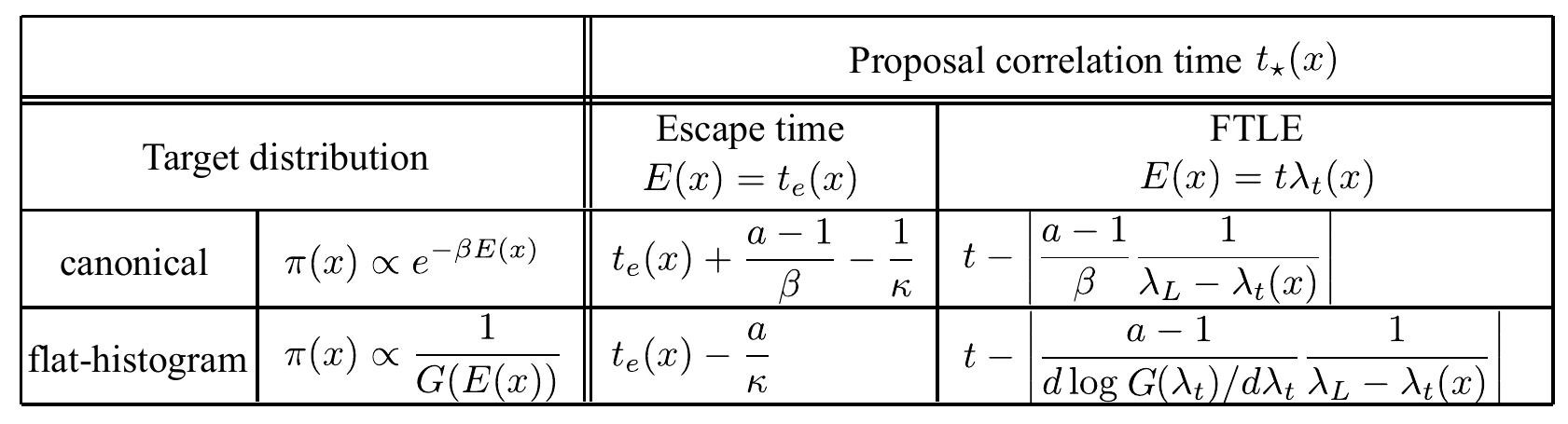}
\caption{
The four $\tstar$ derived for the two problems (escape time and FTLE as the observable) and two target distributions (canonical and flat-histogram).
The four values of $\tstar(\vx)$ reported in the table should be used in $\deltax$ of Eq.~\ref{eq:deltax} and specify the time the proposed trajectory $\vx'$ should stay close to the trajectory $\vx$ for the acceptance to be bounded.
}
\label{tab:summary}
\end{table}

\subsection{Comparison to previous results}

The importance of the proposal distribution has long been emphasized for Monte Carlo methods~\cite{Roberts1997,RobertCasellaBook} and for sampling chaotic systems~\cite{Sweet2001,Dellago2002,Bolhuis2002,Tailleur2007}. The questions that remain open from these works, and that we tackle in our paper, are how the efficiency of the sampling method is related to the proposal distribution and how an efficient proposal can be constructed from assumptions about the system. 
The proposals derived in this paper can be mapped, under appropriate simplifications, to known results from the literature, specifically to the proposals derived Refs.~\cite{Leitao2013,Leitao2014} and the stagger part of the algorithm presented in Ref.~\cite{Sweet2001}. We make this connection explicitly in appendix~\ref{app.simplifications}.
Another example of an algorithm that can be directly analyzed by the framework developed here is the \emph{precision shooting} proposed in Ref.~\cite{Grunwald2008,GrunwaldPhD2009} and used in numerous applications of transition path sampling to chemical reactions~\cite{GrunwaldPhD2009,Rowley2009,Eidelson2012,Schaefer2014}.
The precision shooting method proposes a state $\vx'$ isotropically distanced from $\vx$ by $\deltax$ given by Eq.~\ref{eq:deltax}, with $\tstar(\vx) = \tobs$ where $\tobs$ is the length of the trajectory.
From our results we conclude that this proposal is sub-optimal because it over-correlates the proposed state.
Another application of the results of section~\ref{sec:framework} is in the algorithm \emph{Lyapunov Weighted Dynamics} of Refs.~\cite{Tailleur2007,Laffargue2013a}, which uses a population Monte Carlo algorithm.
As mentioned in these references (see also Ref.~\cite{Wouters2015}), there is a parameter $\varepsilon$ that controls how far new clones $\vx'$ should be distanced from the existing clone $\vx$, and that it should be neither too small nor too large.
This plays a role similar to $\deltax$ in section~\ref{sec:framework} and an optimal $\varepsilon$ should therefore be related to our results.

In comparison to previous sampling methods in dynamical systems, including those mentioned above and others (e.g., ref.~\cite{Sweet2001}), the distinguishing feature of our results is that they provide an explicit connection between the proposal distribution and the acceptance rate. 
This connection, which is typically absent in Monte Carlo methods more generally, is extremely powerful because failures of the algorithm can be related to violations of the hypothesis (about the method and dynamical system) that we used in our derivations.
Such violations should then be understood, and this understanding can then be inserted back in this methodology to generate new methods adapted for that situation.  We hope this process will increase the range of applicability of our framework to other classes of dynamical systems (e.g., non-hyperbolic systems~\cite{Zaslavsky2002,Altmann2013}) and observables $E$. Possible improvements of our results can be obtained considering anistoropic search domains, an idea that has shown to be essential in the case of finding chaotic saddles in systems with more than one positive Lyapunov exponents~\cite{Sala2016}.

\subsection{Discussion}

The proposal distribution is a way of moving in the phase-space stochastically and how to select a new state $\vx'$ from a given state $\vx$ is a general problem in different numerical techniques.
Some of the most successful numerical algorithms in the literature, such as the golden section search or gradient descent, are essentially generic and efficient ways of selecting a new state.
The results in Figs.~\ref{fig:round-trip-lambda}, and ~\ref{fig:lambda_efficiency} show that the proposal distribution strongly influences the computational cost of the different procedures, often irrespectively of the particular sampling procedure (canonical or flat-histogram) and problem (sampling or finding).
This reinforces the notion that the proposal of the new tentative state from the current state is a crucial factor when developing numerical techniques for optimization and numerical integration.
There we can expect that the main insights of this papers, e.g. the direct connection between fundamental properties of chaotic systems and the optimal proposal distribution, to be useful also for other problems (e.g., to the optimization problem in which one is interested in maximizing or minimizing the observable).

The development of a numerical algorithm requires compromising between how fast it solves a particular problem, and how it is able to solve different problems.
One interesting aspect of the algorithms (i.e. proposal distributions) introduced in this paper is that even though they can be made very specific (e.g. propose with the FTLE of the trajectory, Eq.~\ref{eq:deltax}), they can also be made more general (e.g. propose using the Lyapunov exponent of the system, or the power-law proposal distribution, that does not use any specific information about the system).
That is, more specificity requires more information (the FTLE of the trajectory) and makes the algorithm more efficient, and less information (only the Lyapunov of the system) makes the algorithm less specific, but also less efficient.
This demonstrated capability of this methodology shows how it is not only useful to study a particular system on which some information about it is known, but also useful to situations on which less is known about the system.
This does not mean that the methods apply to all problems, as there are important classes of chaotic systems on which some of the assumptions used in section~\ref{sec:framework} are violated.
For example, in non-hyperbolic systems~\cite{Zaslavsky2002,Altmann2013} trajectories may remain correlated for a long time, which implies that one cannot assume that after $\tstar$ the trajectories are independent.
Nevertheless, because the framework was outlined in the form of adding known information about the system, it is possible that improved insights about a class of chaotic systems can be translated to a faster algorithm.

One advantage of the methodology presented here is that it is not restricted to specific observables $E(\vx)$. In principle, it can been used to construct proposal distributions to sample rare states in different observables $E$, $E(\vx)=t \lambda_t$, and $E(\vx) = t_e(\vx)$.
While it remains unclear what is the precise class of observables for which our methodology allows to construct an efficient proposal distribution, the different derivations of the proposal distribution do provide insights on the observables for which a proposal distribution could be constructed from properties of the system.
As argued at the end of Sec~\ref{sec:tstar_ftle}, observables computed as an average along the trajectories are similar to the FTLE and therefore the proposal distribution, derived in Eq.~\ref{eq:tstar_energy} should lead to efficient algorithms in these cases.

Altogether, our results reveal a fascinating interplay between the chaotic nature of some non-linear systems and the numerical techniques available to study rare events. It reinforces the idea that an efficient proposal \emph{requires information about the system} (e.g., our derivation of the proposal distribution used the fact that ``trajectories diverge exponentially'' and that ``the escape time function is a fractal-like function''). The analysis of this interplay allows to both better understand these systems and better understand these numerical techniques.
This understanding opens perspectives to develop better techniques to numerical study rare trajectories and extreme events in non-linear systems more generally.

\newpage

\section{Appendices}
\subsection{Appendix 1: Dynamical Systems}
\label{app:maps}

In this appendix we describe the different dynamical systems that we use throughout the paper.

\subsubsection{Skewed Tent Map}

A paradigmatic example of a strongly chaotic system is the tent map~\cite{OttBook}, defined on $\Omega = [0,1]$ by
\begin{equation}
F(x) = \begin{cases}
a x & \text{ for } x\leq 1/a\\
b(1 - x) & \text{ for } x > 1/a
\end{cases}
\label{eq:tent_map}
\end{equation}
where $a>1$ is a constant and $b \equiv a/(a-1)$
This map contains the main features of a chaotic system: it has a positive Lyapunov exponent ($\lambda_L = \frac{a \log(b) + b \log(a)}{a+b}$) and a positive measure.
The finite-time Lyapunov exponent is given by 
\begin{equation}
\tobs \lambda_{\tobs}(x) = i(x) \log a + \left(\tobs - i(x) \right) \log b \ \ .
\label{eq:lambda_tent_map}
\end{equation}
where $i(x)$ is the number of times $x_t \in [0,1/a]$.
Its distribution of the finite-time Lyapunov exponent for $\uni(\vx) = 1$, $P(E) = G(E)$, can be computed analytically and is a binomial,

\begin{equation}
\begin{split}
G(E) = & \equiv \int_0^1 \delta(E - \tobs \lambda_{\tobs}(\vx)) {\bf d}\vx \\
 & = \sum_{i=0}^{\tobs-1} \delta_{E,\lambda_{\tobs}(i)\tobs} \binom{\tobs}{i} \frac{1}{a^i} \left(1 - \frac{1}{a} \right)^{\tobs - i} \ \ .
\end{split}
\label{eq:rho_lambda_tent_map}
\end{equation}

\subsubsection{Standard map}

The standard map considered here is defined by $\vx \equiv (p, \theta) \in \Omega=[0,1]\times [0,1]$ that evolves in time according to
\begin{equation}
\vF(p, \theta) = \begin{cases}
p + K/(2\pi) \sin(2\pi \theta) \mod 1\\
\theta + p + K/(2\pi) \sin(2\pi \theta) \mod 1\\
\end{cases} \ \ .
\label{eq:standard_map}
\end{equation}
We focus on the parameter $K=6$ which leads to a phase-space with no visible KAM islands.
We also consider the leaked (open) version this map by introducing a hole into the system at $\exitset=[0.1, 0.1]$.

\subsubsection{Skewed Open Tent Map}\label{sssec.tent}

The paradigmatic example of a strongly chaotic open system is the open tent map~\cite{LaiTamasBook}, defined on $\Omega = [0,1]$ by
\begin{equation}
F(x) = \begin{cases}
ax & \text{ for } x\leq b/(a+b)\\
b(1 - x) & \text{ for } x > b/(a+b)
\end{cases}
\label{eq:open_tent}
\end{equation}
where $a>1$ and $b>a/(a-1)$.
The state exits the system when it leaves the unit interval, i.e. $\exitset = \mathbb{R} - \Omega$.
This map contains the main features of an open chaotic system: it has a positive Lyapunov exponent ($\lambda_L = \frac{a \log(b) + b \log(a)}{a+b}$), an exponential decay of the escape time distribution,
\begin{equation}
P(t_e) = \kappa e^{-\kappa t_e} \ \ ,
\label{eq:open_tent_distribution}
\end{equation}
with $\kappa = - \log(1/a + 1/b)$, and a conditionally invariant measure that is fractal with a (non-integer) fractal dimension $D_0$ given implicitly by
\begin{equation}
a^{-D_0} + b^{-D_0} = 1 \ \ .
\label{eq:fractal_eq}
\end{equation}

\subsubsection{Coupled Open H\'enon Map}\label{sec:ncoupled}

As a generic example of a high-dimensional strongly chaotic open system, we consider a set of $d$ coupled H\'enon maps on a ring, defined by a state $\vx = (x_1, y_1, ..., x_{d/2}, y_{d/2}) \in \Omega = \mathbb{R}^{d}$ where each individual map $(x_i, y_i)$ evolves according to
\begin{equation}
\left(\begin{array}{c}
x_{i}\\
y_{i}
\end{array}\right) = 
\left(\begin{array}{c}
A_i - x_i^2 + B y_i + k(x _i - x_{i+1})\\
x_i
\end{array}\right),
\label{eq:nhenon_map}
\end{equation}
for $i = 1,...,d/2$, $d/2 + 1 \equiv 1$, and with parameters $k = 0.4$, $B = 0.3$, $A_1 = 3$ (if $d>1$), $A_{d/2} = 5$, and $A_{i} = A_1 + (A_{d/2} - A_1)(i-1)/(d/2-1)$.
This choice of parameters ensures that a chaotic map is obtained in the $d=2$ case and corresponds to the map studied in Ref.~\cite{Sweet2001} for $d=4$.
The constraining region is $\Gamma = [-4,4]^{d}$ because it covers the chaotic saddle of the system, and $\exitset = \Omega - \Gamma$, i.e. the trajectory leaves the system if the absolute value of any of the coordinates is higher than 4.
The escape function $E(\vx) = t_e(\vx)$ for $d=4$ is represented in figure~\ref{fig:fractal_landscape}.

\subsection{Appendix 2: Efficiency of the uniform proposal}
\label{app:acceptance_derivation}
Here we show that the choice of sampling distribution does not necessarily decrease the scaling of the variance of the estimator.
Our goal is to compute the average acceptance rate for a given escape time $t_e$ in the canonical ensemble with a uniform proposal distribution, $g(\vx'|\vx)=1/|\Gamma|$.
The acceptance is given by $a(\vx'|\vx) = \min\{1,\exp(-\beta (t_e(\vx')-t_e(\vx))\}$ where $\beta < 0$ is used to reach higher $E(\vx) = t_e(\vx)$.
The acceptance of a state $\vx$ is given by 
\begin{equation}
a(\vx) = \int_\Gamma {\bf d}\vx' a(\vx'|\vx) g(\vx'|\vx) \ \ .
\label{eq:acceptance_simple}
\end{equation}
Because $g(\vx'|\vx)=1/|\Gamma|$ and $\pi$ only depends on $t_e(\vx)$, $a(\vx)$ does not depend on $\vx$, only on $t_e$: $a(\vx) = a(t_e(\vx))$.
The average acceptance rate at a given $t_e$, $A(t_e) \equiv \Exp{a(\vx)|t_e}$, is given by
\begin{equation}
A(t_e) = \frac{1}{m(t_e)}\int_\Gamma {\bf d}\vx \delta(t_e - t_e(\vx)) e^{-\beta t_e(\vx)} a(\vx) \ \ .
\end{equation}
Because $a(\vx)$ only depends on $t_e$, it can be pulled out of the integral, and thus $A(t_e) = a(t_e)$.
Taking into account that $P(t_e')= \kappa \exp(- t_e' \kappa)$, the acceptance rate can be computed analytically by integrating Eq.~\ref{eq:acceptance_simple} and leads to 
\begin{equation}
A(t_e) = \frac{e^{-\kappa t_e} \beta + e^{t_e \beta} \kappa}{\beta + \kappa} \ \ .
\end{equation}
This shows that the acceptance rate decays exponentially with increasing $t_e$ (recall that $\beta<0$).
Because low acceptance implies that the random walk stays on the same state for a long time, this leads to an exponential increase of the autocorrelation time $T(E)$ and therefore an increase of the variance in Eq.~\ref{eq:variance3}.
This same argument applies to a flat-histogram simulation, where $\p(\vx) \propto \exp(\kappa t_e(\vx))$.

\subsection{Appendix 3: Simplified proposals}
\label{app.simplifications}

This section presents approximations that can be used to simplify both the implementation time and the computational cost of the proposals derived in Sec.~\ref{sec:framework}.
These approximations often reduce the general proposal derived in Sec.~\ref{sec:framework} to particular proposals already found in the literature, and therefore explains such proposals in this wider context.


\subsubsection{Propose with the Lyapunov exponent in open systems}
\label{sec:lyapunov_open_systems}

 Eq.~\ref{eq:deltax} requires computing $\lambda_{t_e(\vx)}(\vx)$, even though the main interest is in $t_e(\vx)$.
This calculation requires using a numerical algorithm or multiply a product of matrixes~\cite{ChaosBook}, both of which have an associated computational cost.
A simplification to this proposal is to approximate $\lambda_{t_e(\vx)}(\vx)$ by the maximum of the distribution of FTLE with finite-time $t_e$, $\lambda_L(t_e)$,
\begin{equation}
\lambda_{t_e(\vx)}(\vx) \approx \lambda_L(t_e(\vx)) \ \ .
\end{equation}
This approximation is valid as long as $\lambda_{t_e}(\vx)$ is not on the tails of the distribution of FTLE with finite-time $t_e$, $P(\lambda_{t_e})$.
A sampling distribution that only depends on $t_e$, $\p(\vx) = \p(t_e(\vx))$, guarantees that states with the same $t_e$ are equally sampled, $P(\vx|t_e) = \uni(\vx)$, and therefore $P(\lambda_{t_e}(\vx)|t_e) = P(\lambda_{t_e})$.
Furthermore, the approximation of using the maximum of the distribution holds because the tails of $P(\lambda_{t_e})$ decay exponentially with increasing $t_e$ (see sec.~\ref{sec:ftle}).
Under these approximations, Eq.~\ref{eq:deltax} can be simplified to
\begin{equation}
\delta_x(\vx) = \Delta e^{-\lambda_L(t_e(\vx)) \tstar(\vx)} \ \ .
\label{eq:deltax_simplified}
\end{equation}
Furthermore, $\lambda_L(t_e)$ converges to the Lyapunov exponent of the system, $\lambda_L$, with increasing $t_e$.
Therefore, a further simplification is to use the Lyapunov exponent of the system instead of $\lambda_L(t_e)$ in Eq.~\ref{eq:deltax_simplified},
\begin{equation}
\delta_x(\vx) = \Delta e^{-\lambda_L \tstar(\vx)} \ \ .
\label{eq:deltax_lyapunov_step}
\end{equation}
Furthermore, the $\tstar(\vx)$ we derived in Eq.~\ref{eq:tstar_hyperbolic}, in the flat-histogram ensemble and with $\log G(t_e) \propto -\kappa t_e$, can written as
\begin{equation}
\tstar(\vx) = t_e(\vx) - \frac{a}{\kappa}
\end{equation}
Defining the constant $\delta_0 \equiv \Delta e^{-a \lambda_L/\kappa}$, we can write 
\begin{equation}
\delta_x(\vx) = \delta_0 e^{-\lambda_L t_e(\vx)} \ \ .
\label{eq:deltax_lyapunov}
\end{equation}
This equation is exactly the proposal derived in Ref.~\cite{Leitao2013}, and shows that the proposal derived in Sec.~\ref{sec:derivation_te} generalises the proposal in Ref.~\cite{Leitao2013}.

\subsubsection{Adaptively estimate the Lyapunov exponent}

Using the proposal distribution with $\lambda_L$ requires a priori knowledge of it, which typically is not available.
This difficulty resembles the same problem that flat-histogram simulations have: $G(E)$ is required, but it is typically unknown a priori.
This analogy motivates a Monte Carlo procedure that on the fly computes $\delta_x(t)$ that scales with $\lambda_L$.

Consider an hypothetical simulation with an isotropic proposal distribution (Eq.~\ref{eq:isotropic_proposal}) with 
\begin{equation}
\deltax = \sigma(t_e(\vx)) \ \ ,
\label{eq:deltax_adaptive_sigma}
\end{equation}
where $\sigma(t)$ is initially set to be $\sigma(t) = 1$ for every $t$.
Consider also that the simulation reached a state $\vx$ with a high escape time (e.g. $t_e = t_e(\vx) = 10/\kappa$).
A proposed state, $\vx' = \vx + \hat{\vh}\sigma(t_e)$, will most likely have a much lower escape time (e.g. $t_e(\vx') = 1/\kappa$).
From Eq.~\ref{eq:deltax} and Eq.~\ref{eq:tstar_hyperbolic}, this indicates that $\sigma(t_e)$ is much higher than the ``correct'' proposal, $\delta_x(\vx)$, and therefore it should be reduced in the next proposal.
The opposite is also true: when $\sigma(t_e)$ is much smaller than $\delta_x(\vx)$, $t_e(\vx') = t_e(\vx)$ and it should be increased.
This hypothetical simulation suggests that, in the same spirit as the Wang-Landau algorithm to approximate the density $P(t_e)$, there is the possibility to approximate $\delta_x(\vx)$ using an update scheme that can be inserted in the Metropolis-Hastings algorithm, and that is given by the same algorithm as the Wang-Landau (see sec.~\ref{sec:wang-landau}), but instead of updating $P_{WL}(t)$, it updates also $\sigma(t)$~\cite{Leitao2013}:
\begin{equation}
\sigma(t_e) = \begin{cases}
\sigma(t_e) f & \text{ for } t_e(\vx') = t_e \\
\sigma(t_e)/f & \text{ for } t_e(\vx') < t_e \ \ .
\label{eq:sigma_refinement}
\end{cases}
\end{equation}
This update scheme generalises the Wang-Landau procedure to the proposal distribution.  It is expected to converge to a function $\sigma(t)$ that decays exponentially with the Lyapunov exponent of the system, and a proposal distribution with a constant acceptance rate in a flat-histogram simulation. It was extensively tested in different systems (tent map, full chaotic standard map with a leak, Coupled H\'enon map with different $D$s), see Refs.~\cite{Leitao2013,Sala2016,LeitaoPhD}.

\subsubsection{Power-law proposal distribution}
\label{sec:power-law-proposal}

\begin{figure}[!t]
\centering
\includegraphics[width=\columnwidth]{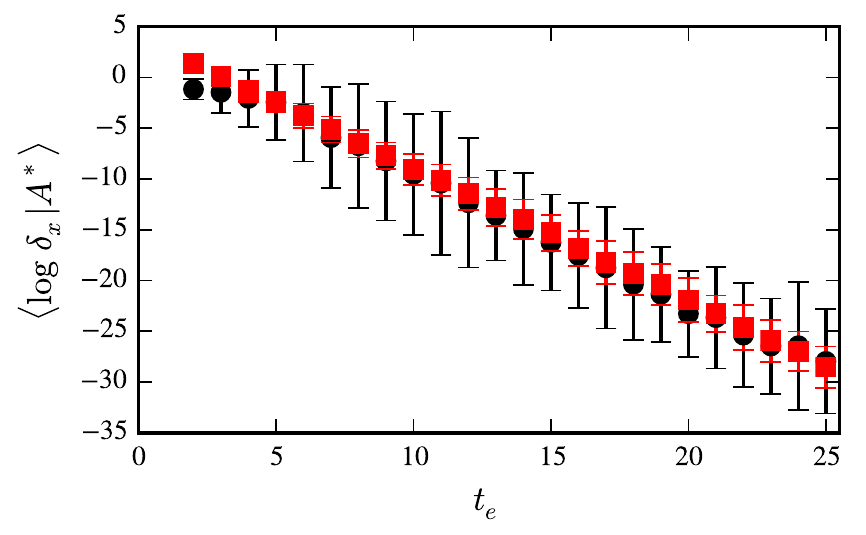}
\caption{
The relevant proposals of the the power-law proposal are those on which the scale is given by $\deltax$ in Eq.~\ref{eq:deltax_lyapunov}.
The x axis is the escape time; the y axis is the best estimator of $\Exp{\log \delta_x | A^*, t_e}$ (black dots, $2\sigma$) over $4\times10^5$ samples obtained from a flat-histogram simulation with the power-law proposal given by Eq.~\ref{eq:power-law}, for each escape time $t_e$ and conditional to on acceptance $A^* = \{\varepsilon<a(\vx'|\vx)<1-\varepsilon \}$ with $\varepsilon = 0.1$.
The power-law proposal distribution samples all scales, but the scales suitable for Metropolis-Hastings depend on $t_e$ as $\exp(-\lambda t_e)$, as expected from the results of Sec.~\ref{sec:lyapunov_open_systems}.
This simulation was made on the open tent map, Eq.~\ref{eq:open_tent}, with $a=3$ and $b=5$.
The best estimate of $-\lambda_{t_e}(\vx) t_e + \log(\Delta)$ with $\Delta = 50$ corresponds to the red line ($2\sigma$). The correspondence of the two curves indicates that the scale of the power-law proposal whose acceptance rate is bounded corresponds to the scale given by $\deltax$.
The parameters used in the power-law proposal were $\delta_{\max}=1$, $\delta_{\min}=2^{-40}$.
}
\label{fig:log_delta_wl_sy}
\end{figure}

The proposal distributions derived in the previous sections requires some knowledge about the state and the system: $\lambda_{\tobs}(\vx)$, $t_e(\vx)$ (in open systems), $\lambda_L$ of the system, and, in some situations, $G(E(\vx))$.
Another alternative to avoid computing $\deltax$ is to consider a proposal on which the time $\tstar$ that the two trajectories remain together is not imposed by a given $\tstar(\vx)$, but that is a uniformly random variable between $[0, \tobs]$ (FTLE) or $[0, t_e(\vx)]$ (open systems) that is generated on each proposal.
Some values of $\tstar$ will be far from the optimal $\tstar(\vx)$ and the corresponding $\vx'$ will be rejected or it will be too close from $\vx$, but others $\tstar$ will still be close from the optimal $\tstar(\vx)$ and therefore useful.

Having a uniformly distributed correlation $\tstar$ still requires computing $\deltax$ in Eq.~\ref{eq:deltax}, which requires $\lambda_{\tobs}(\vx)$.
In the case $\lambda_t(\vx)$ is unknown (e.g. in open systems one could be only interested in the escape time and therefore not compute $\lambda_t(\vx)$), one may further approximate it by an uniform distribution between two extremes.
Because the product of two uniformly distributed random variables is also uniformly distributed, this leads to a proposal distribution where $\deltax$ is given by $\exp(-U(a,b))$ where $a$ and $b$ are free parameters.
By standard transformation of variables, this leads to a scale $\deltax$ that is power-law distributed and given by 
\begin{equation}
P(\delta_x|\vx) = P(\delta_x) = \frac{1}{\delta_x} \frac{1}{s_{\max} - s_{\min}} \in [\delta_{\min}, \delta_{\max}]\ \ ,
\label{eq:power-law}
\end{equation}
where $\delta_{\min}$ and $\delta_{\max} \approx |\Gamma|$ are two free parameters.
The $\delta$ in $\vx' = \vx + \vh \delta$ is an half-normal distribution with a scale $\delta_x$, but since this scale is now power-law distributed, it is no longer necessary to use the half-normal distribution altogether; instead, it is possible to just use the power-law proposal distribution where $\delta$ is drawn from $P(\delta_x)$ in Eq.~\ref{eq:power-law}, i.e. $|\vx'-\vx|$ is power-law distributed according to Eq.~\ref{eq:power-law}.
Without the half-normal distribution the proposal distribution no longer depends on $\vx$ and therefore $g(\vx'|\vx)/g(\vx|\vx') = 1$.

The stagger part of the algorithm of Ref.~\cite{Sweet2001} proposes exactly with a scale given by Eq.~\ref{eq:power-law} and the argumentation above explains why the proposal distribution used in Ref.~\cite{Sweet2001} to find states with high-escape time $t_e$ is reported to work well: it is a proposal distribution that proposes $\vx'$ correlated with $\vx$ with a correlation $\tstar$ that is uniformly distributed, which eventually proposes $\vx'$ with the optimal correlation $\tstar(\vx)$.
To confirm this explanation, let us consider a flat-histogram simulation with a power-law proposal distribution on the open tent map and consider the measurement of $\Exp{\log \delta_x | A^*, t_e}$, where $A^*$ is the condition $\varepsilon<a(\vx'|\vx)<1-\varepsilon$ (of bounded acceptance).
Under the above argumentation, the scale $\log \delta_x$ that contributes to a bounded acceptance is given by $-\lambda_{t_e} t_e$, per Eq.~\ref{eq:deltax_lyapunov}.
This is confirmed by numerical simulation, shown in figure~\ref{fig:log_delta_wl_sy}, and was obtained also for the problem of finding rare states, as reported in Ref.~\cite{Sala2016}.
This result, combined with the derivation of $\tstar(\vx)$, explains the success of the proposal (the stagger) used in Ref.~\cite{Sweet2001} from basic notions of chaotic systems and numerical methods.

\subsection{Appendix 4: Algorithmic description}
\label{app.algorithm}

It is useful to summarise the different proposals in an algorithmic form so they can be easily implemented (see Ref.~\cite{github} for our codes).
In all cases, the proposal requires the current state of the random walk, denoted by $\vx$.

\subsubsection{FTLE in closed systems}

\begin{enumerate}
\item Generate a unitary vector $\hat{\vdelta}$ in $D$ dimensions
\item Compute $\tstar = \tobs - \frac{\log(a)}{d\log \p(E)/dE(E)}\frac{1}{|\lambda_L - \lambda_{\tobs}(\vx)|}$, Eq.~\ref{eq:tstar_lyapunov}.
\item Compute $\delta_x = \Delta \exp(-\tstar \lambda_{\tobs}(\vx))$, Eq.~\ref{eq:deltax}
\item Generate a random number $\delta$ from a normal distribution with mean 0 and variance $\delta_x^2$
\item Make $\vx' = \vx + \hat{\vdelta} |\delta|$
\end{enumerate}
where $\Delta \in \mathbb{R}$ is free parameter (e.g. $\Delta = 0.1$) and $a$ is the chosen average acceptance (e.g. $0.5$).
For example, in a canonic ensemble with parameter $\beta$, $d\log \p(E)/dE(E) = \beta$ and therefore $\tstar = t_e(\vx) - \frac{\log(a)}{\beta}\frac{1}{|\lambda_L - \lambda_{\tobs}(\vx)|}$. The value of $\lambda_L$ can be estimated using e.g. the first samples of the random walk.
In the flat-histogram ensemble, $\lambda_L$ is the maximum of $G(\lambda_{\tobs})$ and $\log \p(E)/dE(E) = \log G(E)/dE(E)$, or by an approximation of it, e.g. using $G_{WL}$ of the Wang-Landau algorithm.

\subsubsection{Open systems}
\label{sec:algorithm_open_systems}

\begin{enumerate}
\item Generate a unitary vector $\hat{\vdelta}$ in $D$ dimensions
\item Compute $\tstar = t_e(\vx) - 1/\kappa - \frac{a - 1}{d\log \pi/dt_e}$, Eq.~\ref{eq:tstar_hyperbolic}
\item Compute $\delta_x = \Delta \exp(-\tstar \lambda_{\tobs}(\vx))$, Eq.~\ref{eq:deltax}
\item Generate a random number $\delta$ from a normal distribution with mean 0 and variance $\delta_x^2$
\item Make $\vx' = \vx + \hat{\vdelta} |\delta|$
\end{enumerate}
where $\delta_0$ is a free parameter (e.g. $\delta_0=0.1$).
Both $\lambda_{t_e}(\vx)$ and $t_e(\vx)$ are required by the proposal and both can be computed during the same evolution of the system:
$t_e(\vx)$ is the time until the trajectory enters the exit region $\Lambda$, $\lambda_{t_e}(\vx)$ is the FTLE of this trajectory.

\paragraph{Acknowledgements.} JCL was funded by FCT (Portugal), Grant NO. SFRH/BD/90050/2012. EGA thanks J. Wouters and T. T\'el for helpful discussions.

\bibliographystyle{apsrev4-1}
\bibliography{library}

\end{document}